\documentclass[aps,prd,showpacs]{revtex4}
\usepackage[latin9]{inputenc}
\usepackage{color}
\usepackage{amsmath}
\usepackage{amssymb}
\usepackage{esint}

\makeatletter
\@ifundefined{textcolor}{}
{%
 \definecolor{BLACK}{gray}{0}
 \definecolor{WHITE}{gray}{1}
 \definecolor{RED}{rgb}{1,0,0}
 \definecolor{GREEN}{rgb}{0,1,0}
 \definecolor{BLUE}{rgb}{0,0,1}
 \definecolor{CYAN}{cmyk}{1,0,0,0}
 \definecolor{MAGENTA}{cmyk}{0,1,0,0}
 \definecolor{YELLOW}{cmyk}{0,0,1,0}
}

\usepackage{amsfonts}\usepackage{bm}\usepackage[dvips]{epsfig}\usepackage{float}\usepackage{dcolumn}\usepackage{ifpdf}\usepackage{dcolumn}\usepackage{multirow}\usepackage{stmaryrd}

\makeatother

\begin{document}

\title{The general form of the coupled Horndeski Lagrangian that allows
cosmological scaling solutions}

\author{Adalto R. Gomes}

\affiliation{Departamento de F\'{i}sica, Universidade Federal do Maranhão (UFMA)
Campus Universitário do Bacanga, 65085-580, São Lu\'{i}s, Maranhão,
Brasil}

\author{Luca Amendola}

\affiliation{Institut für Theoretische Physik, Universität Heidelberg, Philosophenweg
16, D-69120 Heidelberg, Germany.}
\begin{abstract}
We consider the general scalar field Horndeski Lagrangian coupled
to matter. Within this class of models, we present two results\textbf{
}that are independent of the particular form of the model.\textbf{
}First, we show that in a Friedmann-Robertson-Walker metric the Horndeski
Lagrangian coincides with the pressure of the scalar field. Second,
we employ the previous result to identify\textbf{ }the most general
form of the Lagrangian\textbf{ }that allows for cosmological scaling
solutions, i.e. solutions where the ratio of matter to field density
and the equation of state remain constant. Scaling solutions of this
kind may help solving the coincidence problem since in this case the
presently observed ratio of matter to dark energy does not depend
on initial conditions, but rather on the theoretical parameters. 
\end{abstract}
\maketitle

\section{Introduction}


\global\long\def\jcap{JCAP}
 The discovery of cosmic acceleration \cite{Perlmutter_etal_1999,Riess_etal_1998}
has generated many attempts at finding suitable explanations beyond
the classical cosmological constant. Many of these are based on scalar
fields because they provide a relatively simple way of adding degrees
of freedom into the matter Lagrangian with appealing properties: weak
clustering, tunable equation of state, isotropy. In fact, scalar fields
are also the simplest and most investigated models for the other period
of acceleration -- inflation.

The simple quintessence scalar field Lagrangian \cite{Wetterich_1988,Ratra_Peebles_1988,Copeland:1997et,Ferreira_Joyce_1998,1998PhRvL..80.1582C}
has been progressively expanded by including terms coupled to gravity
\cite{Wetterich_1995,amendola2000,Baccigalupi_Matarrese_Perrotta_2000,2006JCAP...12..020A}
and terms that are general functions of the kinetic energy {\cite{kessence}}.
Building on pioneering results, some authors realized recently that
the most general scalar field Lagrangian that produces second order
equations of motion is the so-called Horndeski Lagrangian \cite{Horndeski:1974,Deffayet:2011gz,Kobayashi:2011nu},
a model that includes four arbitrary functions of the scalar field
and its kinetic energy. Models that expand beyond Horndeski have also
been proposed \cite{Gleyzes:2013ooa,2014PhRvD..89f4046Z}. This Lagrangian
is a form of modified gravity since, in general, it changes the Poisson
equation, the lensing equation, and the gravitational wave equation.

The number of free functions in the Horndeski Lagrangian makes an
exhaustive study very complicate. Several classes of models that exhibit
special properties have been already investigated and shown to produce
succesful models of dark energy (see e.g. \cite{2012PhRvD..85j4040C}).
In a previous paper \cite{2014JCAP...03..041G} we posed the question
of whether it was possible to find some general property of the Horndeski
Lagrangian without solving the equations of motion. In particular,
we proposed to identify which classes of Lagrangian contain the so-called
scaling solutions, defined by the property that the energy density
of matter and scalar field scale in the same way with time, so that
their ratio remains constant. A second condition that has also been
employed to simplify the treatment is that the field equation of state
remains constant during the scaling trajectory. Scaling solutions
are particularly interesting because one can hope to employ them to
avoid the problem of the coincidence between the present matter and
dark energy densities, i.e. the fact that today the two density fractions
$\Omega_{m},\Omega_{\phi}$ are very similar. In fact, while this
coincidence occurs only today for a cosmological constant model and
for all the models in which matter and dark energy scale with time
in a different way, and therefore depends in a critical way on the
initial conditions, in scaling solutions the ``coincidence'' depends
only the choice of parameters and, if the solution is stable, this
can remain valid forever. It is important to remark that we are not
demanding that the theory contains \emph{only} scaling solutions,
but that scaling solutions are allowed: in general they will contain
other non-scaling solutions. Whether these models can successfully
explain the entire cosmological sequence of radiation, matter and
accelerated eras is still to be seen \cite{Kscal}.

The first example of scaling model is a simple uncoupled scalar field
with an exponential potential \cite{Copeland:1997et,Ferreira_Joyce_1998}.
However, in this case no\textbf{\textcolor{blue}{{} }}acceleration is
possible during the scaling regime. Next, one can couple the scalar
field and the matter component (or equivalently couple field and gravity)
\cite{Wetterich_1995,amendola2000}. Several interesting properties
of this kind of scaling solutions have been studied in the past, as
for instance a similar coupling to neutrinos \cite{Amendola_Baldi_Wetterich_2008}
and the behavior of perturbations \cite{Amendola:2001rc}, and more
recently, with multiple dark matter models \cite{Baldi_2012a,2015arXiv150205922V}.

A generalization of scaling models has been realized in Ref. \cite{pt}
(see also \cite{ts}). They found in fact that the most general Lagrangian
without gravity coupling that contains scaling solutions must have
the form 
\begin{equation}
S=\int d^{4}x\sqrt{-g}\biggl[\frac{1}{2}R+K(\phi,X)\biggr]+S_{m}(\phi,\psi_{i},g_{\mu\nu})\label{action-1}
\end{equation}
with 
\begin{equation}
K(\phi,X)=Xg(Xe^{\lambda\phi}),
\end{equation}
where $X=-\frac{1}{2}\nabla_{\mu}\phi\nabla^{\mu}\phi$,\, $g$ an
arbitrary function and $\lambda$ a constant. $S_{m}$ is the action
for the matter fields $\psi_{i}$, which also depends generally on
the scalar field $\phi$. The same form applies if the field has a
constant coupling to gravity. In Ref. \cite{Amendola:2006qi} this
result has been extended to variable couplings.

In Ref. \cite{2014JCAP...03..041G} we performed another step towards
extending this result to the entire Horndeski Lagrangian. We studied
in fact a Lagrangian of type \cite{Pujolas:2011he,Deffayet:2010qz,Kobayashi:2011nu}
\begin{equation}
S=\int d^{4}x\sqrt{-g}\biggl[\frac{1}{2}R+K(\phi,X)-G_{3}(\phi,X)\nabla_{\mu}\nabla^{\mu}\phi\biggr]+S_{m}(\phi,\psi_{i},g_{\mu\nu})\label{action-2}
\end{equation}
denoted KGB model in \cite{Deffayet:2010qz}. The new term containing
$G_{3}$ produces new second order terms in the equation of motion.
The addition of the term linear in $\Box\phi\equiv\nabla_{\mu}\nabla^{\mu}\phi$
introduced several new features and enlarged considerably the class
of models that allow for accelerated scaling solutions. This paper
is devoted to completing our programme by extending the search for
scaling Lagrangians to the entire Horndeski class. Unfortunately the
mathematics required to achieve this extension is exceedingly tedious.
The conclusion is however quite simple to state: we obtain\textbf{
}the most general form of Horndeski Lagrangian that contains scaling
solutions\textbf{ }among their Friedmann-Lemaitre-Robertson-Walker
(FLRW) solutions.

To achieve this generalization we need  an intermediate result that
is interesting on its own, namely, the equivalence of the Lagrangian
with the pressure of the scalar field. This is well known to occur
with the term $K(\phi,X)$ \cite{2000PhRvD..62b3511C,ArmendarizPicon:2000dh}
and also when including $G_{3}$ \cite{Deffayet:2010qz}. Here we
prove it valid for the entire Lagrangian, at least when the metric
is restricted to a flat FLRW.

\section{Horndeski Lagrangian and equations of motion}

We consider the Horndeski action 
\begin{equation}
S=\int d^{4}x\sqrt{-g}\mathcal{L}_{H}+S_{m}(\phi,\psi_{i},g_{\mu\nu})\label{action}
\end{equation}
where $\mathcal{L}_{H}$ is the Horndeski Lagrangian, which is the
most general Lagrangian which has second order equations of motion,
defined as $\mathcal{L}_{H}=\mathcal{L}_{2}+\mathcal{L}_{3}+\mathcal{L}_{4}+\mathcal{L}_{5}$,
with 
\begin{eqnarray}
\mathcal{L}_{2} & = & K(\phi,X),\\
\mathcal{L}_{3} & = & -G_{3}(\phi,X)\Box\phi,\\
\mathcal{L}_{4} & = & G_{4}(\phi,X)R+G_{4,X}[(\Box\phi)^{2}-(\nabla_{\mu}\nabla_{\nu}\phi)(\nabla^{\mu}\nabla^{\nu}\phi)]\\
\mathcal{L}_{5} & = & G_{5}(\phi,X)G_{\mu\nu}(\nabla^{\mu}\nabla^{\nu}\phi)-\frac{1}{6}G_{5,X}[(\Box\phi)^{3}-3(\Box\phi)(\nabla_{\mu}\nabla_{\nu}\phi)(\nabla^{\mu}\nabla^{\nu}\phi)\nonumber \\
 &  & +2(\nabla^{\mu}\nabla_{\alpha}\phi)(\nabla^{\alpha}\nabla_{\beta}\phi)(\nabla^{\beta}\nabla_{\mu}\phi)]
\end{eqnarray}
and $\phi$ is a scalar field, $X=-\frac{1}{2}\nabla_{\mu}\phi\nabla^{\mu}\phi$
and $R$ is the Ricci scalar. We consider that there is only one type
of pressureless matter of energy density $\rho_{m}=-T_{0}^{0}$, in
the Einstein frame, where the energy-momentum tensor is defined by
\begin{equation}
T_{\mu\nu}=-\frac{2}{\sqrt{-g}}\frac{\delta S_{m}}{\delta g^{\mu\nu}}.
\end{equation}
In this frame matter is directly coupled to the scalar field through
the function $Q(\phi)$, where 
\begin{equation}
Q=-\frac{1}{\rho_{m}\sqrt{-g}}\frac{\delta S_{m}}{\delta\phi}.\label{Qdef}
\end{equation}

We consider a FLRW flat metric with $ds^{2}=-dt^{2}+\mathcal{A}^{2}(t)d\textbf{x}^{2}$,
where $\mathcal{A}(t)$ is the scale factor and the scalar field depends
only on $t$. In this case we have $X=\dot{\phi}^{2}/2$, $\dot{X}=\dot{\phi}\ddot{\phi}$,
$\nabla_{\mu}X\nabla^{\mu}\phi=-2X\ddot{\phi}$, where dot means derivative
with respect to the cosmic time $t$. Varying the action with respect
to $g_{\mu\nu}$, and defining $H=\dot{\mathcal{A}}/{\mathcal{A}}$,
one gets \cite{DeFelice:2011bh} 
\begin{eqnarray}
\sum_{i=2}^{5}\mathcal{E}_{i} & = & -\rho_{m},\label{eq_gmunu_a}\\
\sum_{i=2}^{5}\mathcal{P}_{i} & = & 0,\label{eq_gmunu_b}
\end{eqnarray}
where 
\begin{eqnarray}
\mathcal{E}_{2} & = & 2XK_{,X}-K,\\
\mathcal{E}_{3} & = & 6X\dot{\phi}HG_{3,X}-2XG_{3,\phi},\\
\mathcal{E}_{4} & = & -6H^{2}G_{4}+24H^{2}X(G_{4,X}+XG_{4,XX})-12HX\dot{\phi}G_{4,\phi X}-6H\dot{\phi}G_{4,\phi},\\
\mathcal{E}_{5} & = & 2H^{3}X\dot{\phi}(5G_{5,X}+2XG_{5,XX})-6H^{2}X(3G_{5,\phi}+2XG_{5,\phi,X}).
\end{eqnarray}
and 
\begin{eqnarray}
\mathcal{P}_{2} & = & K,\\
\mathcal{P}_{3} & = & -2X(G_{3,\phi}+\ddot{\phi}G_{3,X}),\\
\mathcal{P}_{4} & = & 2(3H^{2}+2\dot{H})G_{4}-12H^{2}XG_{4,X}-4H\dot{X}G_{4,X}-8\dot{H}XG_{4,X}-8HX\dot{X}G_{4,XX}\nonumber \\
 &  & +2(\ddot{\phi}+2H\dot{\phi})G_{4,\phi}+4XG_{4,\phi\phi}+4X(\ddot{\phi}-2H\dot{\phi})G_{4,\phi X},\\
\mathcal{P}_{5} & = & -2X(2H^{3}\dot{\phi}+2H\dot{H}\dot{\phi}+3H^{2}\ddot{\phi})G_{5,X}-4H^{2}X^{2}\ddot{\phi}G_{5,XX}\nonumber \\
 &  & +4HX(\dot{X}-HX)G_{5,\phi X}+2[2(\dot{H}X+H\dot{X})+3H^{2}X]G_{5,\phi}+4HX\dot{\phi}G_{5,\phi\phi}.
\end{eqnarray}

Varying the action with respect to $\phi$, one gets 
\begin{equation}
\frac{1}{\mathcal{A}^{3}}\frac{d}{dt}(a^{3}J)=\mathcal{P}-\rho_{m}Q,
\end{equation}
where 
\begin{eqnarray}
J & \equiv & \dot{\phi}K_{,X}+6HXG_{3,X}-2\dot{\phi}G_{3,\phi}+6H^{2}\dot{\phi}(G_{4,X}+2XG_{4,XX})-12HXG_{4,\phi X}\nonumber \\
 &  & +2H^{3}X(3G_{5,X}+2XG_{5,XX})-6H^{2}\dot{\phi}(G_{5,\phi}+XG_{5,\phi X})\\
\mathcal{P} & \equiv & K_{,\phi}-2X(G_{3,\phi\phi}+\ddot{\phi}G_{3,\phi X})+6(2H^{2}+\dot{H})G_{4,\phi}+6H(\dot{X}+2HX)G_{4,\phi X}\nonumber \\
 &  & -6H^{2}XG_{5,\phi\phi}+2H^{3}X\dot{\phi}G_{5,\phi X}
\end{eqnarray}

Now, in order to confront these models with SNIa observations, we
isolate from the complete action a term corresponding to the Einstein-Hilbert
one. Then the action is rewritten as 
\begin{equation}
S=\int d^{4}x\sqrt{-g}\biggl(\frac{1}{2}R+\mathcal{L}\biggr)+S_{m}(\phi,\psi_{i},g_{\mu\nu})\label{action_R}
\end{equation}
with 
\begin{eqnarray}
\mathcal{L} & = & K(\phi,X)-G_{3}(\phi,X)\Box\phi+\biggl(G_{4}(\phi,X)-\frac{1}{2}\biggr)R+G_{4,X}[(\Box\phi)^{2}-(\nabla_{\mu}\nabla_{\nu}\phi)(\nabla^{\mu}\nabla^{\nu}\phi)]\nonumber \\
 &  & +G_{5}(\phi,X)G_{\mu\nu}(\nabla^{\mu}\nabla^{\nu}\phi)-\frac{1}{6}G_{5,X}[(\Box\phi)^{3}-3(\Box\phi)(\nabla_{\mu}\nabla_{\nu}\phi)(\nabla^{\mu}\nabla^{\nu}\phi)\nonumber \\
 &  & +2(\nabla^{\mu}\nabla_{\alpha}\phi)(\nabla^{\alpha}\nabla_{\beta}\phi)(\nabla^{\beta}\nabla_{\mu}\phi)
\end{eqnarray}
and we write the Einstein equations as \cite{tsujikawa_DMDE} 
\begin{equation}
H^{2}=\frac{1}{3}(\rho_{\phi}+\rho_{m})\label{feq}
\end{equation}
and 
\begin{equation}
-2\dot{H}=\rho_{m}+\rho_{\phi}+p.\label{feq2}
\end{equation}

Comparing Eqs. (\ref{feq}), (\ref{feq2}) with Eqs. (\ref{eq_gmunu_a}),
(\ref{eq_gmunu_b}), we arrive to useful definitions for the energy
density ($\rho_{DE}$) and pressure ($p$) of dark energy: 
\begin{eqnarray}
\rho_{\phi} & \equiv & \sum_{i=2}^{5}\mathcal{E}_{i}+3H^{2},\\
p & \equiv & \sum_{i=2}^{5}\mathcal{P}_{i}-(3H^{2}+2\dot{H}),
\end{eqnarray}
or also 
\begin{eqnarray}
\rho_{\phi} & \equiv & 2XK_{,X}-K+6X\dot{\phi}HG_{3,X}-2XG_{3,\phi}+3H^{2}(1-2G_{4})\nonumber \\
 &  & +24H^{2}X(G_{4,X}+XG_{4,XX})-12HX\dot{\phi}G_{4,\phi X}-6H\dot{\phi}G_{4,\phi}\nonumber \\
 &  & +2H^{3}X\dot{\phi}(5G_{5,X}+2XG_{5,XX})-6H^{2}X(3G_{5,\phi}+2XG_{5,\phi,X}),\\
p & \equiv & K-2X(G_{3,\phi}+\ddot{\phi}G_{3,X})-(3H^{2}+2\dot{H})(1-2G_{4})-12H^{2}XG_{4,X}\nonumber \\
 &  & -4H\dot{X}G_{4,X}-8\dot{H}XG_{4,X}-8HX\dot{X}G_{4,XX}+2(\ddot{\phi}+2H\dot{\phi})G_{4,\phi}\nonumber \\
 &  & +4XG_{4,\phi\phi}+4X(\ddot{\phi}-2H\dot{\phi})G_{4,\phi X}\nonumber \\
 &  & -2X(2H^{3}\dot{\phi}+2H\dot{H}\dot{\phi}+3H^{2}\ddot{\phi})G_{5,X}-4H^{2}X^{2}\ddot{\phi}G_{5,XX}\nonumber \\
 &  & +4HX(\dot{X}-HX)G_{5,\phi X}+2[2(\dot{H}X+H\dot{X})+3H^{2}X]G_{5,\phi}+4HX\dot{\phi}G_{5,\phi\phi}.
\end{eqnarray}
Defining 
\begin{equation}
\Omega_{\phi}=\frac{\rho_{\phi}}{3H^{2}},\,\,\,\Omega_{m}=\frac{\rho_{m}}{3H^{2}}\label{Omegaphi}
\end{equation}
we can rewrite the Friedman equation (Eq. (\ref{feq})) as 
\begin{equation}
\Omega_{\phi}+\Omega_{m}=1.\label{sumOmega}
\end{equation}
Now we introduce the $e$-folding time $N=\log a$, so that $d/dt=Hd/dN$.
Then the equation of motion for the scalar field $\phi$ and matter
can be written as (see also \cite{DeFelice:2011bh}) 
\begin{align}
\frac{d\rho_{\phi}}{dN}+3(1+w_{\phi})\rho_{\phi} & =-\rho_{m}Q\frac{d\phi}{dN}\label{rhophi2}\\
\frac{d\rho_{m}}{dN}+3\rho_{m} & =\rho_{m}Q\frac{d\phi}{dN},\label{rhom2}
\end{align}
where $w_{\phi}=p/\rho_{\phi}$.


\section{Scaling Solutions}

The condition $\Omega_{\phi}/\Omega_{m}$ constant defines scaling
solutions. This is equivalent to $\rho_{\phi}/\rho_{m}$ constant,
or to 
\begin{equation}
\frac{d\log\rho_{\phi}}{dN}=\frac{d\log\rho_{m}}{dN}\label{scaling}
\end{equation}
Also, from Eq. (\ref{sumOmega}) we get that $\Omega_{\phi}$ is a
constant. We also assume that on scaling solutions the equation of
state parameter $w_{\phi}$ is a constant \cite{ts}. Subtracting
both Eqs. (\ref{rhophi2}) and (\ref{rhom2}) and using Eq. (\ref{scaling})
we get 
\begin{equation}
\frac{d\phi}{dN}=-\frac{3\Omega_{\phi}w_{\phi}}{Q}\propto\frac{1}{Q}.\label{dphidN}
\end{equation}
Back to Eqs. (\ref{rhophi2}) and (\ref{rhom2}) we get 
\begin{equation}
\frac{d\log\rho_{\phi}}{dN}=\frac{d\log\rho_{m}}{dN}=-3(1+w_{eff}),
\end{equation}
where $w_{eff}=w_{\phi}\Omega_{\phi}$. Now, from $w_{\phi}$ constant,
we have 
\begin{equation}
\frac{d\log p}{dN}=-3(1+w_{eff})\label{dlogp_dn}
\end{equation}
At this point we need a crucial statement, namely that the Lagrangian
$\mathcal{L}$ is equivalent to the pressure of the scalar field,
up to boundary terms. The demonstration that indeed this is true for
the entire Horndeski Lagrangian is rather long and tedious and we
moved it to the Appendix C. For our purposes, it is enough to show
this to be true for a flat FLRW metric, since we are looking for scaling
solutions only on such a metric. We conjecture that $\mathcal{L}=p$
for any metric; this indeed can easily be shown to be the case for
the $K$ and the $G_{3}$ terms and we will present the general proof
elsewhere.

From the equality $\mathcal{L}=p$ we have then 
\begin{eqnarray}
p & = & K-G_{3}\Box\phi+\biggl(-\frac{1}{2}+G_{4}\biggr)R+G_{4,X}[(\Box\phi)^{2}-\boxtimes\phi]\nonumber \\
 &  & +G_{5}(\phi,X)(\boxbar\phi)-\frac{1}{6}G_{5,X}[(\Box\phi)^{3}-3(\Box\phi)\boxtimes\phi+2(\boxdot\phi)
\end{eqnarray}
with 
\begin{eqnarray}
\boxtimes\phi & \equiv & (\nabla_{\mu}\nabla_{\nu}\phi)(\nabla^{\mu}\nabla^{\nu}\phi)\\
\boxbar\phi & \equiv & G_{\mu\nu}(\nabla^{\mu}\nabla^{\nu}\phi)\\
\boxdot\phi & \equiv & (\nabla^{\mu}\nabla_{\alpha}\phi)(\nabla^{\alpha}\nabla_{\beta}\phi)(\nabla^{\beta}\nabla^{\mu}\phi)
\end{eqnarray}
For FLRW metric, we have $R=6\dot{H}+12H^{2}$. Then, using Einstein
equations, we get 
\begin{equation}
R=\biggl(\frac{1}{w_{\phi}\Omega_{\phi}}-3\biggr)p.
\end{equation}
That is, for scaling solutions we have $R=\tilde{c}p$, with $\tilde{c}$
a constant.

Now defining 
\begin{equation}
f(\phi,X)\equiv\biggl[1+\biggl(\frac{1}{2}-G_{4}(\phi,X)\biggr)\tilde{c}\biggr],\label{f_def}
\end{equation}
we can write the pressure of dark energy as 
\begin{eqnarray}
p & = & \frac{1}{f(\phi,X)}\biggl[K(\phi,X)-G_{3}(\phi,X)\Box\phi+G_{4,X}[(\Box\phi)^{2}-\boxtimes\phi]\nonumber \\
 &  & +G_{5}(\phi,X)(\boxbar\phi)-\frac{1}{6}G_{5,X}[(\Box\phi)^{3}-3(\Box\phi)\boxtimes\phi+2(\boxdot\phi)\biggr]\label{eq:p_finv}
\end{eqnarray}
That is, $p=p(X,\Box{\phi},\boxtimes\phi,\boxbar\phi,\boxdot\phi,\phi)$

Now we want to find a generalized ``master equation'' for $p=p(X,\Box{\phi},\boxtimes\phi,\boxbar\phi,\boxdot\phi,\phi)$.
Eq. (\ref{dlogp_dn}) gives 
\begin{eqnarray}
 &  & \frac{\partial\log p}{\partial\phi}\frac{d\phi}{dN}+\frac{\partial\log p}{\partial\log X}\frac{d\log X}{dN}+\frac{\partial\log p}{\partial\log\Box\phi}\frac{d\log\Box\phi}{dN}+\frac{\partial\log p}{\partial\log\boxtimes\phi}\frac{d\log\boxtimes\phi}{dN}\nonumber \\
 &  & +\frac{\partial\log p}{\partial\log\boxbar\phi}\frac{d\log\boxbar\phi}{dN}+\frac{\partial\log p}{\partial\log\boxdot\phi}\frac{d\log\boxdot\phi}{dN}=-3(1+w_{eff}).\label{master}
\end{eqnarray}
Now we know that

\begin{eqnarray*}
\frac{d\phi}{dN} & = & -\frac{3\Omega_{\phi}w_{\phi}}{Q}=3(1+w_{eff})\frac{1}{\lambda Q},
\end{eqnarray*}
with 
\begin{equation}
\lambda=-\frac{1+w_{eff}}{w_{eff}}.\label{lambda}
\end{equation}
We need the partial derivatives ${d\log X}/{dN}$, ${d\log\Box\phi}/{dN}$,
${d\log\boxtimes\phi}/{dN}$, ${d\log\boxbar\phi}/{dN}$ and ${d\log\boxdot\phi}/{dN}$
that are obtained as follows (for details, see Appendix C):

\subsection{${d\log X}/{dN}$}

From the definition of $X$ and Eq. (\ref{dphidN}) we have 
\begin{equation}
X=\frac{1}{2}\dot{\phi}^{2}=\frac{H^{2}}{2}\biggl(\frac{d\phi}{dN}\biggr)^{2}\propto\frac{H^{2}}{Q^{2}}=\frac{\rho_{\phi}}{3\Omega_{\phi}}\frac{1}{Q^{2}}\propto\frac{p}{Q^{2}},
\end{equation}
and then 
\begin{eqnarray}
\frac{d\log X}{dN} & = & \frac{d\log p}{dN}-2\frac{d\log Q}{dN}\nonumber \\
 & = & -3(1+w_{\mathrm{eff}})-\frac{2}{Q}\frac{dQ}{d\phi}\frac{d\phi}{dN}\\
 & = & -3(1+w_{\mathrm{eff}})-\frac{2}{Q}\frac{dQ}{d\phi}\left(-\frac{3\Omega_{\phi}w_{\phi}}{Q}\right)\\
 & = & -3(1+w_{\mathrm{eff}})\left(1+\frac{2}{\lambda Q^{2}}\frac{dQ}{d\phi}\right)
\end{eqnarray}

\subsection{${d\log\Box\phi}/{dN}$}

We start with (for details here and forthe next terms, see Appendices
A and B) 
\begin{equation}
\Box\phi=-3H\dot{\phi}-\ddot{\phi}\label{Boxphi0}
\end{equation}
Now, from Eq. (\ref{dphidN}) this can be rewritten as 
\begin{equation}
\Box\phi=\frac{3}{2}(1-w_{eff})\frac{p}{Q}\left(1-\frac{2}{\lambda}\frac{1+w_{\mathrm{eff}}}{1-w_{\mathrm{eff}}}\frac{1}{Q^{2}}\frac{dQ}{d\phi}\right),\label{Boxphi}
\end{equation}
So far we put no restrictions on the coupling function $Q$. However
we find that the analysis is very simplified if we assume 
\begin{equation}
\frac{1}{Q^{2}}\frac{dQ}{d\phi}=c=const.\label{eqQ}
\end{equation}
This restricts the coupling to be 
\begin{equation}
Q(\phi)=\frac{1}{-c\phi+c_{2}},\label{Qsol}
\end{equation}
with $c_{2}$ constant. From now we we will consider this restriction
in $Q(\phi)$. From Eq. (\ref{Boxphi}) we have then 
\begin{equation}
\log\Box\phi=\log p-\log Q+const.,
\end{equation}
which gives 
\begin{equation}
\frac{\partial\log\Box\phi}{\partial N}=-3(1+w_{eff})\biggl(1+\frac{1}{\lambda Q^{2}}\frac{dQ}{d\phi}\biggr).
\end{equation}

\subsection{${d\log\boxtimes\phi}/{dN}$ }

We start with 
\begin{equation}
\boxtimes\phi=(\ddot{\phi})^{2}+3H^{2}(\dot{\phi})^{2}.
\end{equation}
This results in 
\begin{equation}
\frac{\partial\log\boxtimes\phi}{\partial N}=-3(1+w_{eff})\biggl(2+\frac{2}{\lambda Q^{2}}\frac{dQ}{d\phi}\biggr).
\end{equation}

\subsection{${d\log\boxbar\phi}/{dN}$}

We start with 
\begin{equation}
\boxbar\phi=-3H^{2}\ddot{\phi}-6H\dot{H}\dot{\phi}-9H^{3}\dot{\phi}.
\end{equation}
This results in 
\begin{equation}
\frac{\partial\log\boxbar\phi}{\partial N}=-3(1+w_{eff})\biggl(2+\frac{1}{\lambda Q^{2}}\frac{dQ}{d\phi}\biggr).
\end{equation}

\subsection{${d\log\boxdot\phi}/{dN}$}

We start with 
\begin{equation}
\boxdot\phi=-\ddot{\phi}^{3}-3H^{3}\dot{\phi}^{3}.
\end{equation}
This results in 
\begin{equation}
\frac{\partial\log\boxdot\phi}{\partial N}=-3(1+w_{eff})\biggl(3+\frac{3}{\lambda Q^{2}}\frac{dQ}{d\phi}\biggr).
\end{equation}
Finally, Eq. (\ref{master}) becomes 
\begin{eqnarray}
 &  & \biggl(1+\frac{2}{\lambda Q^{2}}\frac{dQ}{d\phi}\biggr)\frac{\partial\log p}{\partial\log X}+\biggl(1+\frac{1}{\lambda Q^{2}}\frac{dQ}{d\phi}\biggr)\frac{\partial\log p}{\partial\log\Box\phi}+\biggl(2+\frac{2}{\lambda Q^{2}}\frac{dQ}{d\phi}\biggr)\frac{\partial\log p}{\partial\log\boxtimes\phi}\nonumber \\
 &  & +\biggl(2+\frac{1}{\lambda Q^{2}}\frac{dQ}{d\phi}\biggr)\frac{\partial\log p}{\partial\log\boxbar\phi}+\biggl(3+\frac{3}{\lambda Q^{2}}\frac{dQ}{d\phi}\biggr)\frac{\partial\log p}{\partial\log\boxdot\phi}-\frac{1}{\lambda Q}\frac{\partial\log p}{\partial\phi}=1.\label{master2}
\end{eqnarray}
As expected, the master equation reduces to the one obtained in Ref.
\cite{Amendola:2006qi} when $G_{3}(\phi,X)=G_{5}(\phi,X)=0$ and
$G_{4}(\phi,X)=1/2$.


\section{Solutions for the master equation}

Here, after a convenient \textit{Ansatz}, we derive the general solution
for the master equation Eq. (\ref{master2}). Remember, however, that
there are restrictions in the form of $Q(\phi)$, given by Eq. (\ref{Qsol}),
that will be taken into account in due course. We start with Eq. (\ref{master2})
rewritten as 
\begin{eqnarray}
 &  & \biggl(1+\frac{2}{Q}\frac{dQ}{d\psi}\biggr)\frac{\partial\log p}{\partial\log X}+\biggl(1+\frac{1}{Q}\frac{dQ}{d\psi}\biggr)\frac{\partial\log p}{\partial\log\Box\phi}+\biggl(2+\frac{2}{Q}\frac{dQ}{d\psi}\biggr)\frac{\partial\log p}{\partial\log\boxtimes\phi}\nonumber \\
 &  & +\biggl(2+\frac{1}{Q}\frac{dQ}{d\psi}\biggr)\frac{\partial\log p}{\partial\log\boxbar\phi}+\biggl(3+\frac{3}{Q}\frac{dQ}{d\psi}\biggr)\frac{\partial\log p}{\partial\log\boxdot\phi}-\frac{\partial\log p}{\partial\psi}=1,\label{master3}
\end{eqnarray}
where 
\begin{equation}
\psi=\int_{\phi}du[\lambda Q(u)].\label{psi-def}
\end{equation}

Now set 
\begin{equation}
p=XQ^{2}(\phi)\tilde{G}(X,\Box\phi,\boxtimes\phi,\boxbar\phi,\boxdot\phi,\phi).\label{p-tildeg-Q}
\end{equation}
where $\tilde{G}$ is an arbitray function of its argument. Then for
$\tilde{G}\neq0$ we obtain 
\begin{eqnarray}
 &  & \biggl(1+\frac{2}{Q}\frac{dQ}{d\psi}\biggr)X\frac{\partial\tilde{G}}{\partial X}+\biggl(1+\frac{1}{Q}\frac{dQ}{d\psi}\biggr)\Box\phi\frac{\partial\tilde{G}}{\partial\Box\phi}+\biggl(2+\frac{2}{Q}\frac{dQ}{d\psi}\biggr)(\boxtimes\phi)\frac{\partial\tilde{G}}{\partial(\boxtimes\phi)}\nonumber \\
 &  & +\biggl(2+\frac{1}{Q}\frac{dQ}{d\psi}\biggr)(\boxbar\phi)\frac{\partial\tilde{G}}{\partial(\boxbar\phi)}+\biggl(3+\frac{3}{Q}\frac{dQ}{d\psi}\biggr)(\boxdot\phi)\frac{\partial\tilde{G}}{\partial(\boxdot\phi)}-\frac{\partial\tilde{G}}{\partial\psi}=0.\label{masterQ}
\end{eqnarray}
where by (\ref{eqQ}) the term $\frac{2}{Q}\frac{dQ}{d\psi}$ is a
constant. This partial differential equation is linear in $\tilde{G}$.
Then the method of separation of variables is justifiable. Here, however,
inspired by the expression given by Eq. (\ref{eq:p_finv}) for $p$,
we look for solutions of $\tilde{G}$ of the form 
\begin{equation}
\tilde{G}(X,\Box\phi,\boxtimes\phi,\boxbar\phi,\boxdot\phi,\phi)=\frac{1}{f(\phi,X)}{\tilde{g}(X,\Box\phi,\boxtimes\phi,\boxbar\phi,\boxdot\phi,\phi)}.
\end{equation}
Then, Eq. (\ref{masterQ}) turns into 
\begin{eqnarray}
 &  & \frac{1}{f}\biggl\{\biggl[\biggl(1+\frac{2}{Q}\frac{dQ}{d\psi}\biggr)X\frac{\partial f}{\partial X}-\frac{\partial f}{\partial\psi}\biggr]\frac{\tilde{g}}{f}\nonumber \\
 &  & -\biggl[\biggl(1+\frac{2}{Q}\frac{dQ}{d\psi}\biggr)X\frac{\partial\tilde{g}}{\partial X}+\biggl(1+\frac{1}{Q}\frac{dQ}{d\psi}\biggr)\Box\phi\frac{\partial\tilde{g}}{\partial\Box\phi}+\biggl(2+\frac{2}{Q}\frac{dQ}{d\psi}\biggr)(\boxtimes\phi)\frac{\partial\tilde{g}}{\partial(\boxtimes\phi)}\nonumber \\
 &  & +\biggl(2+\frac{1}{Q}\frac{dQ}{d\psi}\biggr)(\boxbar\phi)\frac{\partial\tilde{g}}{\partial(\boxbar\phi)}+\biggl(3+\frac{3}{Q}\frac{dQ}{d\psi}\biggr)(\boxdot\phi)\frac{\partial\tilde{g}}{\partial(\boxdot\phi)}-\frac{\partial\tilde{g}}{\partial\psi}\biggr]\biggr\}=0.
\end{eqnarray}
For $f\neq0$ this is equivalent to $(\widehat{O}_{1}f)\tilde{G}-\widehat{O}_{2}\tilde{g}=0,$where
the operators $\widehat{O}_{1}$ and $\widehat{O}_{2}$ are evident
in the equation above. For $G_{4}=1/2$ and $G_{5}=0$ we have $f=1$
and the linear differential equation $\widehat{O}_{2}\tilde{g}=0$
is reduced to a form already solved in our previous paper\cite{2014JCAP...03..041G}
. Now we want to generalize our former result for general $G_{4}$
and $G_{5}$. The simplest choice is to impose that $\tilde{g}$ and
$f$ satisfy separately the linear differential equations: 
\begin{equation}
\widehat{O}_{1}f=0,
\end{equation}
that is 
\begin{equation}
\biggl(1+\frac{2}{Q}\frac{dQ}{d\psi}\biggr)X\frac{\partial f}{\partial X}-\frac{\partial f}{\partial\psi}=0
\end{equation}
with solution (known for the case where $G_{2}\neq0$, $G_{4}=1/2$
and $G_{3}=G_{5}=0)$. 
\begin{equation}
f(\phi,X)=g_{2}(XQ^{2}e^{\psi}),
\end{equation}
where $g_{2}$ is a general function. Now we turn to the equation
for $\tilde{g}$: 
\begin{equation}
\widehat{O}_{2}g=0,
\end{equation}
that is 
\begin{eqnarray}
 &  & \biggl(1+\frac{2}{Q}\frac{dQ}{d\psi}\biggr)X\frac{\partial\tilde{g}}{\partial X}+\biggl(1+\frac{1}{Q}\frac{dQ}{d\psi}\biggr)\Box\phi\frac{\partial\tilde{g}}{\partial\Box\phi}+\biggl(2+\frac{2}{Q}\frac{dQ}{d\psi}\biggr)(\boxtimes\phi)\frac{\partial\tilde{g}}{\partial(\boxtimes\phi)}\nonumber \\
 &  & +\biggl(2+\frac{1}{Q}\frac{dQ}{d\psi}\biggr)(\boxbar\phi)\frac{\partial\tilde{g}}{\partial(\boxbar\phi)}+\biggl(3+\frac{3}{Q}\frac{dQ}{d\psi}\biggr)(\boxdot\phi)\frac{\partial\tilde{g}}{\partial(\boxdot\phi)}-\frac{\partial\tilde{g}}{\partial\psi}=0
\end{eqnarray}

We do not want the general solution of this linear differential equation.
Indeed, guided by the form of $p$, we look for a solution of the
form (all other possibilites lead to trivial constant solutions or
are not compatible with the form of the Lagrangian). 
\begin{eqnarray}
\tilde{g} & = & g_{b}(h_{b})+g_{c}(h_{c})+g_{d}(h_{d})+g_{\ell}(h_{\ell})+g_{e}(h_{e})+g_{j}(h_{j})+g_{m}(h_{m})+g_{n}(h_{n})+g_{p}(h_{p})\nonumber \\
 &  & +g_{q}(h_{q})+g_{r}(h_{r})+g_{s}(h_{s})+g_{t}(h_{t})+g_{u}(h_{u})+g_{z}(h_{v})+g_{z}(h_{z}),\label{ga-gd-Q}
\end{eqnarray}
where $g_{b},g_{c},g_{d},...$ are arbitrary functions and 
\begin{eqnarray}
h_{b}(X,\psi) & = & f_{1b}(X)f_{3b}(\psi),\label{hbQ}\\
h_{c}(X,\Box\phi) & = & f_{1c}(X)f_{2c}(\Box\phi)\label{hcQ}\\
h_{d}(\Box\phi,\psi) & = & f_{2d}(\Box\phi)f_{3d}(\psi).\label{hdQ}\\
h_{\ell}(X,\Box\phi,\psi) & = & f_{1\ell}(X)f_{2\ell}(\Box\phi)f_{3\ell}(\psi)\label{hlQ}\\
h_{e}(X,\boxtimes\phi) & = & f_{1e}(X)f_{4e}(\boxtimes\phi)\label{heQ}\\
h_{j}(\boxtimes\phi,\psi) & = & f_{3j}(\psi)f_{4j}(\boxtimes\phi)\label{hjQ}\\
h_{m}(X,\boxtimes\phi,\psi) & = & f_{1m}(X)f_{3m}(\psi)f_{4m}(\boxtimes\phi)\label{hmQ}\\
h_{n}(\psi,\boxbar\phi) & = & f_{2n}(\psi)f_{5n}(\boxbar\phi)\label{hnQ}\\
h_{p}(X,\boxbar\phi) & = & f_{1p}(X)f_{5p}(\boxbar\phi)\label{hpQ}\\
h_{q}(X,\boxbar\phi,\psi) & = & f_{1q}(X)f_{3q}(\psi)f_{5q}(\boxbar\phi)\label{hqQ}\\
h_{r}(X,\boxdot\phi) & = & f_{1r}(X)f_{6r}(\boxdot\phi)\label{hrQ}\\
h_{s}(\psi,\boxdot\phi) & = & f_{2s}(\psi)f_{6s}(\boxdot\phi)\label{hsQ}\\
h_{t}(X,\boxdot\phi,\psi) & = & f_{1t}(X)f_{3t}(\psi)f_{6t}(\boxdot\phi)\label{htQ}\\
h_{u}(X,\Box\phi,\boxtimes\phi) & = & f_{1u}(X)f_{2u}(\Box\phi)f_{4u}(\boxtimes\phi)\label{huQ}\\
h_{v}(\psi,\Box\phi,\boxtimes\phi) & = & f_{3v}(\psi)f_{2v}(\Box\phi)f_{4v}(\boxtimes\phi)\label{hvQ}\\
h_{z}(\psi,X,\Box\phi,\boxtimes\phi) & = & f_{1z}(X)f_{3z}(\psi)f_{2z}(\Box\phi)f_{4z}(\boxtimes\phi)\label{hzQ}
\end{eqnarray}
In the following we will consider separately each possibility. We
will be guided by the general form given by Eq. (\ref{eq:p_finv})
for $p$, which fixes the maximum order of the factors $\Box\phi,\boxtimes\phi,\boxbar\phi$
and $(\boxdot\phi)$.

\subsection{$g_{b}(h_{b}(X,\psi))$}

Eq. (\ref{masterQ}) gives 
\begin{equation}
\frac{dg}{dh_{b}}\biggl[\biggl(1+\frac{2}{Q}\frac{dQ}{d\psi}\biggr)\frac{d\log f_{1b}}{d\log X}-\frac{1}{f_{3b}}\frac{df_{3b}}{d\psi}\biggr]=0,
\end{equation}
which gives $f_{1b}=X^{\alpha}$ and $f_{3b}=e^{\alpha\psi}Q^{2\alpha}$.
Then Eq. (\ref{hbQ}) gives 
\begin{equation}
h_{b}=(XQ^{2}(\phi)e^{\psi})^{\alpha}
\end{equation}
and therefore 
\begin{equation}
g(h_{b})=g(XQ^{2}(\phi)e^{\psi}),
\end{equation}
where $g$ is an arbitrary function.

\subsection{$g_{c}(h_{c}(X,\Box\phi))$}

This gives $g_{c}$ constant.

\subsection{$g_{d}(h_{d}(\Box\phi,\psi))$}

Eq. (\ref{masterQ}) gives 
\begin{equation}
\frac{dg_{d}}{dh_{d}}\biggl[\biggl(1+\frac{1}{Q}\frac{dQ}{d\psi}\biggr)\frac{d\log f_{2d}}{d\log\Box\phi}-\frac{1}{f_{3d}}\frac{df_{3d}}{d\psi}\biggr]=0.\label{master-gd-Q}
\end{equation}
which gives $f_{2d}=(\Box\phi)^{\alpha}$ and $f_{3d}=Q^{\alpha}e^{\alpha\psi}$.
Then Eq. (\ref{hdQ}) gives 
\begin{equation}
h_{d}=\biggl((\Box\phi)e^{\psi}Q\biggr)^{\alpha}
\end{equation}
and therefore 
\begin{equation}
g_{d}(h_{d})=g_{d}\biggl((\Box\phi)Q(\phi)e^{\psi}\biggr).
\end{equation}
Now comparing with Eq. (\ref{eq:p_finv}) for $p$, we see that $g_{d}(h_{d})$
must be at most third order in $\Box\phi$. Then we have 
\begin{equation}
g_{d}(h_{d})=d_{1}\biggl((\Box\phi)Q(\phi)e^{\psi}\biggr)+d_{2}\biggl((\Box\phi)Q(\phi)e^{\psi}\biggr)^{2}+d_{3}\biggl((\Box\phi)Q(\phi)e^{\psi}\biggr)^{3},
\end{equation}
with $d_{1},d_{2},d_{3}$ constants.

\subsection{$g_{\ell}(h_{\ell}(X,\Box\phi,\psi))$}

Eq. (\ref{masterQ}) gives 
\begin{equation}
\frac{dg_{\ell}}{dh_{\ell}}\biggl[\biggl(1+\frac{2}{Q}\frac{dQ}{d\psi}\biggr)\frac{1}{f_{1\ell}}\frac{df_{1\ell}}{d\log X}+\biggl(1+\frac{1}{Q}\frac{dQ}{d\psi}\biggr)\frac{1}{f_{2\ell}}\frac{df_{2\ell}}{d\log\Box\phi}-\frac{1}{f_{3\ell}}\frac{df_{3\ell}}{d\psi}\biggr]=0
\end{equation}
which gives $f_{1\ell}=X^{\alpha}$, $f_{2\ell}=(\Box\phi)^{\beta}$
and $f_{3\ell}=e^{(\alpha+\beta)\psi}Q^{2\alpha+\beta}$. Then Eq.
(\ref{hlQ}) gives 
\begin{equation}
h_{\ell}=\biggl[X(\Box\phi)^{\beta/\alpha}e^{(1+\beta/\alpha)\psi}Q^{2+\beta/\alpha}\biggr]^{\alpha}
\end{equation}
and 
\begin{equation}
g_{\ell}(h_{\ell})=g_{\ell}\biggl(X(\Box\phi)^{n}e^{(1+n)\psi}Q^{2+n}\biggr),
\end{equation}
where from Eq. (\ref{eq:p_finv}) for $p$, the exponent $n$ can
be $1$, $2$ or $3$ for terms linear, quadratic or cubic in $(\Box\phi)$.
This leads to 
\begin{equation}
g_{\ell}(h_{\ell})={\ell_{1}}\biggl(X(\Box\phi)e^{2\psi}Q^{3}(\phi)\biggr)+{\ell_{2}}\biggl(X(\Box\phi)^{2}e^{3\psi}Q^{4}(\phi)\biggr)+{\ell_{3}}\biggl(X(\Box\phi)^{3}e^{4\psi}Q^{5}(\phi)\biggr),
\end{equation}
with $\ell_{1},\ell_{2},\ell_{3}$ constants

\subsection{$g_{e}(h_{e}(X,\boxtimes\phi))$}

This gives $g_{e}$ constant.

\subsection{$g_{j}(h_{j}(\boxtimes\phi,\psi))$}

Eq. (\ref{masterQ}) gives 
\begin{equation}
\frac{dg_{j}}{dh_{j}}\biggl[\biggl(2+\frac{2}{Q}\frac{dQ}{d\psi}\biggr)\frac{d\log f_{4j}}{d\log\boxtimes\phi}-\frac{1}{f_{3j}}\frac{df_{3j}}{d\psi}\biggr]=0.\label{master-gj-Q}
\end{equation}
which gives $f_{4j}=(\boxtimes\phi)^{\alpha}$ and $f_{3j}=Q^{2\alpha}e^{2\alpha\psi}$.
Then Eq. (\ref{hjQ}) gives 
\begin{equation}
h_{j}=\biggl((\boxtimes\phi)e^{2\psi}Q^{2}\biggr)^{\alpha}.
\end{equation}
Comparing with Eq. (\ref{eq:p_finv}) for $p$, we see that $g_{j}(h_{j})$
must be at most linear in $\Box\phi$. Then we get 
\begin{equation}
g_{j}(h_{j})={j}\biggl((\boxtimes\phi)Q^{2}e^{2\psi}\biggr),
\end{equation}
with $j$ a constant.

\subsection{$g_{m}(h_{m}(X,\boxtimes\phi,\psi))$}

Eq. (\ref{masterQ}) gives 
\begin{equation}
\frac{dg_{m}}{dh_{m}}\biggl[\biggl(1+\frac{2}{Q}\frac{dQ}{d\psi}\biggr)\frac{1}{f_{1m}}\frac{df_{1m}}{d\log X}+\biggl(2+\frac{2}{Q}\frac{dQ}{d\psi}\biggr)\frac{1}{f_{2m}}\frac{d\log f_{2m}}{d\log\boxtimes\phi}-\frac{1}{f_{3m}}\frac{df_{3m}}{d\psi}\biggr]=0
\end{equation}
which gives $f_{1m}=X^{\alpha}$, $f_{4m}=(\boxtimes\phi)^{\beta}$
and $f_{3m}=e^{(\alpha+2\beta)\psi}Q^{2\alpha+2\beta}$. Then Eq.
(\ref{hmQ}) gives 
\begin{equation}
h_{m}=\biggl[X(\boxtimes\phi)^{\beta/\alpha}e^{(1+2\beta/\alpha)\psi}Q^{2+2\beta/\alpha}\biggr]^{\alpha}.
\end{equation}
Then comparing with Eq. (\ref{eq:p_finv}) we get 
\begin{equation}
g_{m}(h_{m})={m}\biggl(X(\boxtimes\phi)e^{3\psi}Q^{4}\biggr),
\end{equation}
with $m$ a constant.

\subsection{$g_{p}(h_{p}(X,\boxbar\phi))$}

This gives $g_{p}$ constant.

\subsection{$g_{n}(h_{n}(\boxbar\phi,\psi))$}

Eq. (\ref{masterQ}) gives 
\begin{equation}
\frac{dg_{n}}{dh_{n}}\biggl[\biggl(2+\frac{1}{Q}\frac{dQ}{d\psi}\biggr)\frac{d\log f_{5n}}{d\log\boxbar\phi}-\frac{1}{f_{3n}}\frac{df_{3n}}{d\psi}\biggr]=0.
\end{equation}
which gives $f_{5n}=(\boxbar\phi)^{\alpha}$ and $f_{3n}=Q^{\alpha}e^{2\alpha\psi}$.
Then Eq. (\ref{hjQ}) gives 
\begin{equation}
h_{n}=\biggl((\boxbar\phi)e^{2\psi}Q\biggr)^{\alpha}.
\end{equation}
Comparing with Eq. (\ref{eq:p_finv}), we see that $g_{n}(h_{n})$
must be at most linear in $(\boxbar\phi)$.Then we get 
\begin{equation}
g_{n}(h_{n})={n}\biggl((\boxbar\phi)Qe^{2\psi}\biggr),
\end{equation}
with $n$ a constant.

\subsection{$g_{q}(h_{q}(X,\boxbar\phi,\psi))$}

Eq. (\ref{masterQ}) gives 
\begin{equation}
\frac{dg_{q}}{dh_{q}}\biggl[\biggl(1+\frac{2}{Q}\frac{dQ}{d\psi}\biggr)\frac{1}{f_{1q}}\frac{df_{1q}}{d\log X}+\biggl(2+\frac{1}{Q}\frac{dQ}{d\psi}\biggr)\frac{1}{f_{5q}}\frac{d\log f_{5q}}{d\log\boxbar\phi}-\frac{1}{f_{3q}}\frac{df_{3q}}{d\psi}\biggr]=0
\end{equation}
which gives $f_{1q}=X^{\alpha}$, $f_{5q}=(\boxbar\phi)^{\beta}$
and $f_{3q}=e^{(\alpha+2\beta)\psi}Q^{2\alpha+\beta}$. Then Eq. (\ref{hmQ})
gives 
\begin{equation}
h_{q}=\biggl[X(\boxbar\phi)^{\beta/\alpha}e^{(1+2\beta/\alpha)\psi}Q^{2+\beta/\alpha}\biggr]^{\alpha}.
\end{equation}
Then comparing with Eq. (\ref{eq:p_finv}), we get 
\begin{equation}
g_{q}(h_{q})={q}\biggl(X(\boxbar\phi)e^{3\psi}Q^{3}\biggr),
\end{equation}
with $m$ a constant.

\subsection{$g_{r}(h_{r}(X,\boxdot\phi))$}

This gives $g_{r}$ constant.

\subsection{$g_{s}(h_{s}(\boxdot\phi,\psi))$}

Eq. (\ref{masterQ}) gives 
\begin{equation}
\frac{dg_{s}}{dh_{s}}\biggl[\biggl(3+\frac{3}{Q}\frac{dQ}{d\psi}\biggr)\frac{d\log f_{6s}}{d\log\boxdot\phi}-\frac{1}{f_{3s}}\frac{df_{3s}}{d\psi}\biggr]=0.
\end{equation}
which gives $f_{6s}=(\boxdot\phi)^{\alpha}$ and $f_{3n}=Q^{3\alpha}e^{3\alpha\psi}$.
Then Eq. (\ref{hjQ}) gives 
\begin{equation}
h_{s}=\biggl((\boxbar\phi)e^{3\psi}Q^{3}\biggr)^{\alpha}.
\end{equation}
Then comparing with Eq. (\ref{eq:p_finv}), we get 
\begin{equation}
g_{s}(h_{n})={s}\biggl((\boxdot\phi)Q^{3}e^{3\psi}\biggr),
\end{equation}
with $s$ a constant.

\subsection{$g_{t}(h_{t}(X,\boxdot\phi,\psi))$}

Eq. (\ref{masterQ}) gives 
\begin{equation}
\frac{dg_{t}}{dh_{t}}\biggl[\biggl(1+\frac{2}{Q}\frac{dQ}{d\psi}\biggr)\frac{1}{f_{1t}}\frac{df_{1t}}{d\log X}+\biggl(3+\frac{3}{Q}\frac{dQ}{d\psi}\biggr)\frac{1}{f_{6t}}\frac{df_{6t}}{d\log\boxdot\phi}-\frac{1}{f_{3t}}\frac{df_{3t}}{d\psi}\biggr]=0
\end{equation}
which gives $f_{1t}=X^{\alpha}$, $f_{6t}=(\boxdot\phi)^{\beta}$
and $f_{3t}=e^{(\alpha+3\beta)\psi}Q^{2\alpha+3\beta}$. Then Eq.
(\ref{htQ}) gives 
\begin{equation}
h_{t}=\biggl[X(\boxdot\phi)^{\beta/\alpha}e^{(1+3\beta/\alpha)\psi}Q^{2+3\beta/\alpha}\biggr]^{\alpha}.
\end{equation}
Comparing with Eq. (\ref{eq:p_finv}) for $p$, we see that $g_{t}(h_{t})$
must be at most linear in $(\boxbar\phi)$. Then we get 
\begin{equation}
g_{t}(h_{t})={t}\biggl(X(\boxdot\phi)e^{4\psi}Q^{5}\biggr),
\end{equation}
with $t$ a constant.

\subsection{$g_{u}(h_{u}(X,\Box\phi,\boxtimes\phi))$}

This gives $g_{u}$ constant.

\subsection{$g_{v}(h_{v}(\Box\phi,\boxtimes\phi,\psi))$}

Eq. (\ref{masterQ}) gives 
\begin{equation}
\frac{dg_{v}}{dh_{v}}\biggl[\biggl(1+\frac{1}{Q}\frac{dQ}{d\psi}\biggr)\frac{d\log f_{1v}}{d\log X}+\biggl(2+\frac{2}{Q}\frac{dQ}{d\psi}\biggr)\frac{d\log f_{4v}}{d\log\boxtimes\phi}-\frac{1}{f_{3s}}\frac{df_{3v}}{d\psi}\biggr]=0.
\end{equation}
which gives 
\begin{equation}
g_{v}(h_{v})={v}\biggl((\Box\phi)(\boxtimes\phi)Q^{3}e^{3\psi}\biggr),
\end{equation}
with $v$ a constant.

\subsection{$g_{z}(h_{z}(X,\Box\phi,\boxtimes\phi,\psi))$}

Eq. (\ref{masterQ}) gives 
\begin{eqnarray}
 &  & \frac{dg_{z}}{dh_{z}}\biggl[\biggl(1+\frac{2}{Q}\frac{dQ}{d\psi}\biggr)\frac{1}{f_{1z}}\frac{df_{1z}}{d\log X}+\biggl(1+\frac{1}{Q}\frac{dQ}{d\psi}\biggr)\frac{1}{f_{2z}}\frac{df_{2z}}{d\log\Box\phi}\nonumber \\
 &  & +\biggl(2+\frac{2}{Q}\frac{dQ}{d\psi}\biggr)\frac{1}{f_{4z}}\frac{df_{4z}}{d\log\boxtimes\phi}-\frac{1}{f_{3z}}\frac{df_{3z}}{d\psi}\biggr]=0
\end{eqnarray}
which gives $f_{1z}=X^{\alpha}$, $f_{2z}=(\Box\phi)^{\beta}$, $f_{4z}=(\boxtimes\phi)^{\beta}$,
and $f_{3t}=e^{(\alpha+3\beta)\psi}Q^{2\alpha+3\beta}$. Then Eq.
(\ref{htQ}) gives 
\begin{equation}
h_{z}=\biggl[X(\Box\phi\boxtimes\phi)^{\beta/\alpha}e^{(1+3\beta/\alpha)\psi}Q^{2+3\beta/\alpha}\biggr]^{\alpha}.
\end{equation}
Then comparing with Eq. (\ref{eq:p_finv}), we get 
\begin{equation}
g_{t}(h_{z})={z}\biggl(X(\Box\phi\boxtimes\phi)e^{4\psi}Q^{5}\biggr),
\end{equation}
with $z$ a constant.

\subsection{General form for Lagrangian with scaling solution}

From the former results and Eq. (\ref{ga-gd-Q}) we obtain

\begin{eqnarray}
\tilde{g}(X,\Box\phi,\boxtimes\phi,\boxbar\phi,\boxdot\phi,\phi) & = & g(XQ^{2}(\phi)e^{\psi})+d_{1}\biggl((\Box\phi)Q(\phi)e^{\psi}\biggr)+d_{2}\biggl((\Box\phi)Q(\phi)e^{\psi}\biggr)^{2}\nonumber \\
 &  & +d_{3}\biggl((\Box\phi)Q(\phi)e^{\psi}\biggr)^{3}+{\ell_{1}}\biggl(X(\Box\phi)e^{2\psi}Q^{3}(\phi)\biggr)+{\ell_{2}}\biggl(X(\Box\phi)^{2}e^{3\psi}Q^{4}(\phi)\biggr)\nonumber \\
 &  & +{\ell_{3}}\biggl(X(\Box\phi)^{3}e^{4\psi}Q^{5}(\phi)\biggr)+j(Q^{2}(\phi)e^{2\psi}(\boxtimes\phi))+{m}(XQ^{4}(\phi)e^{3\psi}(\boxtimes\phi))\nonumber \\
 &  & +{n}\biggl((\boxbar\phi)Q(\phi)e^{2\psi}\biggr)+{q}\biggl(X(\boxbar\phi)e^{3\psi}Q^{3}\biggr)+{s}\biggl((\boxdot\phi)Q(\phi)^{3}e^{3\psi}\biggr)\nonumber \\
 &  & +{t}\biggl(X(\boxdot\phi)e^{4\psi}Q^{5}\biggr)+{v}\biggl((\Box\phi)(\boxtimes\phi)Q^{3}e^{3\psi}\biggr)\nonumber \\
 &  & +{z}\biggl(X(\Box\phi\boxtimes\phi)e^{4\psi}Q^{5}\biggr).
\end{eqnarray}
We have also obtained 
\begin{equation}
f(\phi,X)=g_{2}(XQ^{2}e^{\psi}),
\end{equation}
Now we need to compare our results and Eq. (\ref{p-tildeg-Q}) , rewritten
as 
\begin{equation}
p=\frac{1}{f(\phi,X)}[XQ^{2}(\phi)\tilde{g}(X,\Box\phi,\boxtimes\phi,\boxbar\phi,\boxdot\phi,\phi)]
\end{equation}
with the expression of $p$ given by Eq. (\ref{eq:p_finv})\textbf{,}
rewritten as: 
\begin{eqnarray}
p & = & \frac{1}{\biggl[1+\biggl(\frac{1}{2}-G_{4}(\phi,X)\biggr)\tilde{c}\biggr]}\biggl[K(\phi,X)-G_{3}(\phi,X)\Box\phi+G_{4,X}[(\Box\phi)^{2}-\boxtimes\phi]\nonumber \\
 &  & +G_{5}(\phi,X)(\boxbar\phi)-\frac{1}{6}G_{5,X}[(\Box\phi)^{3}-3(\Box\phi)\boxtimes\phi+2(\boxdot\phi)\biggr].
\end{eqnarray}
This results in 
\begin{eqnarray}
1+\biggl(\frac{1}{2}-G_{4}(\phi,X)\biggr)\tilde{c} & = & g_{2}(XQ^{2}e^{\psi})\\
K(\phi,X) & = & XQ^{2}g(XQ^{2}(\phi)e^{\psi})\\
-G_{3}(\phi,X)\Box\phi & = & XQ^{2}\biggl[d_{1}\biggl((\Box\phi)Q(\phi)e^{\psi}\biggr)\nonumber \\
 &  & +{\ell_{1}}\biggl(X(\Box\phi)e^{2\psi}Q^{3}(\phi)\biggr)\biggr]\\
G_{4,X}[(\Box\phi)^{2}-\boxtimes\phi] & = & XQ^{2}\biggl[d_{2}\biggl((\Box\phi)Q(\phi)e^{\psi}\biggr)^{2}\nonumber \\
 &  & +{\ell_{2}}\biggl(X(\Box\phi)^{2}e^{3\psi}Q^{4}(\phi)\biggr)\nonumber \\
 &  & +j(Q^{2}(\phi)e^{2\psi}\boxtimes\phi)\nonumber \\
 &  & +{m}(XQ^{4}(\phi)e^{3\psi}\boxtimes\phi)\biggr]\\
G_{5}(\phi,X)(\boxbar\phi) & = & XQ^{2}\biggl[{n}\biggl((\boxbar\phi)Q(\phi)e^{2\psi}\biggr)\nonumber \\
 &  & +{q}\biggl(X(\boxbar\phi)e^{3\psi}Q^{3}\biggr)\biggr]\\
-\frac{1}{6}G_{5,X}[(\Box\phi)^{3}-3(\Box\phi)\boxtimes\phi+2(\boxdot\phi)\biggr] & = & XQ^{2}\biggl[d_{3}\biggl((\Box\phi)Q(\phi)e^{\psi}\biggr)^{3}\nonumber \\
 &  & +{\ell_{3}}\biggl(X(\Box\phi)^{3}e^{4\psi}Q^{5}(\phi)\biggr)\nonumber \\
 &  & +{v}\biggl((\Box\phi)(\boxtimes\phi)Q^{3}e^{3\psi}\biggr)\nonumber \\
 &  & {z}\biggl(X(\Box\phi\boxtimes\phi)e^{4\psi}Q^{5}\biggr)\nonumber \\
 &  & +{s}\biggl((\boxdot\phi)Q(\phi)^{3}e^{3\psi}\biggr)\nonumber \\
 &  & +{t}\biggl(X(\boxdot\phi)e^{4\psi}Q^{5}\biggr)\biggr]
\end{eqnarray}
Now, in each of the former equations from above we must impose relations
between the general constants in order to get a compatible solution
for the general functions. 
\begin{itemize}
\item For $G_{4}$:
\begin{equation}
d_{2}=-j,
\end{equation}
\begin{equation}
\ell_{2}=-m,
\end{equation}

\item For $G_{5}$: 
\begin{eqnarray}
n & = & 0,\\
\ell_{3} & = & 0,\\
t & = & 0,\\
z & = & 0\\
v & = & -3d_{3},\\
s & = & 2d_{3},
\end{eqnarray}

\end{itemize}
Additionally, consistency between expressions for $G_{5}$ and $G_{5,X}$
gives

\begin{eqnarray}
d_{3} & = & -\frac{1}{3}q.
\end{eqnarray}
The above conditions and consistency between expressions for $G_{4}$
and $G_{4,X}$ result in the following expressions for the functions
that compose the Horndeski Lagrangian: 
\begin{eqnarray}
K(\phi,X) & = & XQ^{2}g(XQ^{2}e^{\psi})\\
G_{3}(\phi,X) & = & -d_{1}XQ^{3}e^{\psi}-{\ell_{1}}X^{2}Q^{5}e^{2\psi}\\
G_{4}(\phi,X) & = & h(\phi)+\frac{1}{2}d_{2}X^{2}Q^{4}e^{2\psi}+\frac{1}{3}\ell_{2}Xe^{3\psi}Q^{6}\\
G_{5}(\phi,X) & = & qX^{2}e^{3\psi}Q^{5}.
\end{eqnarray}
where $h(\phi)$ is a general smooth function of $\phi$. Then the
general scaling Horndeski Lagrangian is 
\begin{eqnarray}
\mathcal{L}_{H} & = & XQ^{2}g(XQ^{2}e^{\psi})-[d_{1}XQ^{3}e^{\psi}+{\ell_{1}}X^{2}Q^{5}e^{2\psi}]\Box\phi\nonumber \\
 &  & +\biggl(h(\phi)+\frac{1}{2}d_{2}X^{2}Q^{4}e^{2\psi}+\frac{1}{3}{\ell_{2}}X^{3}e^{3\psi}Q^{6}\biggr)R\nonumber \\
 &  & +\biggl(d_{2}XQ^{4}e^{2\psi}+\ell_{2}X^{2}e^{3\psi}Q^{6}\biggr)[(\Box\phi)^{2}-(\nabla_{\mu}\nabla_{\nu}\phi)(\nabla^{\mu}\nabla^{\nu}\phi)]\nonumber \\
 &  & +qX^{2}e^{3\psi}Q^{5}G_{\mu\nu}\nabla^{\mu}\nabla^{\nu}\phi\nonumber \\
 &  & -\biggl(\frac{1}{6}\biggr)2qXe^{3\psi}Q^{5}[(\Box\phi)^{3}-3(\Box\phi)(\nabla_{\mu}\nabla_{\nu}\phi)(\nabla^{\mu}\nabla^{\nu}\phi)\nonumber \\
 &  & +2(\nabla^{\mu}\nabla_{\alpha}\phi)(\nabla^{\alpha}\nabla_{\beta}\phi)(\nabla^{\beta}\nabla_{\mu}\phi)].\label{L_scal}
\end{eqnarray}
Note that case $d_{2}=\ell_{2}=q=0$ and $h(\phi)=\frac{1}{2}$ recover
the solution we have found in Ref. \cite{2014JCAP...03..041G} for
scaling cosmological solutions in the KGB model. 

Now, in order to ease the comparison with the literature, let us make
the following field redefinitions. First of all take $\psi\to\lambda\psi$.
Then

\begin{equation}
\psi(\phi)=\int_{\phi}duQ(u).\label{psi-def2}
\end{equation}
 Now consider $\phi\to\psi(\phi)$, with $\psi(\phi)$ given by Eq.
(\ref{psi-def2}). This implies 

\begin{eqnarray}
X & \to & X_{\psi}=XQ^{2}(\phi)\\
Q\Box\phi & \to & \Box\psi+2\frac{d\log Q}{d\psi}X_{\psi}\\
Q\nabla^{\mu}\nabla^{\nu}\phi & \to & \nabla^{\mu}\nabla^{\nu}\psi-\frac{d\log Q}{d\psi}\partial^{\mu}\psi\partial^{\nu}\psi\\
Q^{2}\boxtimes\phi & \to & \boxtimes\psi-2\frac{d\log Q}{d\psi}\partial_{\mu}\psi\partial_{\nu}\psi\nabla^{\mu}\nabla^{\nu}\psi+4\biggl(\frac{d\log Q}{d\psi}\biggr)^{2}X_{\psi}^{2}\\
Q^{3}\boxdot\phi & \to & \boxdot\psi-\frac{d\log Q}{d\psi}\biggl(\partial^{\mu}\psi\partial_{\alpha}\psi\nabla^{\alpha}\nabla_{\beta}\psi\nabla^{\beta}\nabla_{\mu}\psi+\nabla^{\mu}\psi\nabla_{\alpha}\psi\partial^{\alpha}\partial_{\beta}\psi\nabla^{\beta}\nabla_{\mu}\psi\nonumber \\
 &  & +\nabla^{\mu}\psi\nabla_{\alpha}\psi\nabla^{\alpha}\nabla_{\beta}\psi\partial^{\beta}\partial_{\mu}\psi\biggr)\nonumber \\
 &  & -\left(\frac{d\log Q}{d\psi}\right)^{2}\biggl(\nabla^{\mu}\nabla_{\alpha}\psi\partial^{\alpha}\psi\partial_{\beta}\psi\partial^{\beta}\psi\partial_{\mu}\psi+\partial^{\mu}\psi\partial_{\alpha}\psi\nabla^{\alpha}\nabla_{\beta}\psi\partial^{\beta}\psi\partial_{\mu}\psi\nonumber \\
 &  & +\partial^{\mu}\psi\partial_{\alpha}\psi\partial^{\alpha}\psi\partial_{\beta}\psi\nabla^{\beta}\nabla_{\mu}\psi\biggr)-\left(\frac{d\log Q}{d\psi}\right)^{3}(8X_{\psi}^{2})\\
\frac{1}{Q^{2}}\frac{dQ}{d\phi} & \to & \frac{d\log Q}{d\psi}
\end{eqnarray}
Then Eq. (\ref{L_scal}) turns into 
\begin{eqnarray}
\mathcal{L}_{H} & = & X_{\psi}g(X_{\psi}e^{\lambda\psi})-[d_{1}X_{\psi}e^{\lambda\psi}+{\ell_{1}}X_{\psi}^{2}e^{2\lambda\psi}]\left(\Box\psi+2\frac{d\ln Q}{d\psi}X_{\psi}\right)\nonumber \\
 &  & +\biggl(h(\psi)+\frac{1}{2}d_{2}X_{\psi}^{2}e^{2\lambda\psi}+\frac{1}{3}{\ell_{2}}X_{\psi}^{3}e^{3\lambda\psi}\biggr)R\nonumber \\
 &  & +\biggl(d_{2}X_{\psi}e^{2\lambda\psi}+\ell_{2}X_{\psi}^{2}e^{3\lambda\psi}\biggr)\left[(\Box\psi)^{2}-\boxtimes\phi+4\frac{d\ln Q}{d\psi}\left(X_{\psi}\Box\psi+\frac{1}{2}\partial^{\mu}\psi\partial_{\nu}\psi\nabla_{\mu}\nabla^{\nu}\psi\right)\right]\nonumber \\
 &  & +qX_{\psi}^{2}e^{3\lambda\psi}G_{\mu\nu}\left(\nabla^{\mu}\nabla^{\nu}\psi-\frac{d\ln Q}{d\psi}\partial^{\mu}\psi\partial^{\nu}\psi\right)\nonumber \\
 &  & -\biggl(\frac{1}{6}\biggr)2qX_{\psi}e^{3\lambda\psi}\biggl\{(\Box\psi)^{3}-3(\Box\psi)\boxtimes\psi+2\boxdot\psi\nonumber \\
 &  & +\frac{d\ln Q}{d\psi}\biggl[6(\Box\psi)^{2}X_{\psi}-6X_{\psi}\boxtimes\psi+6(\Box\psi)\partial_{\mu}\psi\partial_{\nu}\psi\nabla^{\mu}\nabla^{\nu}\psi\nonumber \\
 &  & -2\nabla^{\mu}\nabla_{\alpha}\psi\nabla^{\alpha}\nabla_{\beta}\psi\partial^{\beta}\psi\partial_{\mu}\psi-2\partial^{\mu}\psi\partial_{\alpha}\psi\nabla^{\alpha}\nabla_{\beta}\psi\nabla^{\beta}\nabla{}_{\mu}\psi\nonumber \\
 &  & -2\nabla^{\mu}\nabla_{\alpha}\psi\partial^{\alpha}\psi\partial_{\beta}\psi\nabla^{\beta}\nabla_{\mu}\psi\biggr)\biggr]\nonumber \\
 &  & +\left(\frac{d\log Q}{d\psi}\right)^{2}\biggl[12X_{\psi}\partial_{\mu}\psi\partial_{\nu}\psi\nabla^{\mu}\nabla^{\nu}\psi-2\nabla^{\mu}\nabla_{\alpha}\psi\partial^{\alpha}\psi\partial_{\beta}\psi\partial^{\beta}\psi\partial_{\mu}\psi\nonumber \\
 &  & -2\partial^{\mu}\psi\partial_{\alpha}\psi\nabla^{\alpha}\nabla_{\beta}\psi\partial^{\beta}\psi\partial_{\mu}\psi-2\partial^{\mu}\psi\partial_{\alpha}\psi\partial^{\alpha}\psi\partial_{\beta}\psi\nabla^{\beta}\nabla_{\mu}\psi\biggr]\nonumber \\
 &  & +\left(\frac{d\log Q}{d\psi}\right)^{3}(-8X_{\psi}^{3})\biggr\}.
\end{eqnarray}
It could seem that the Lagrangian has an explicit dependence on the
coupling $Q(\phi)$. However, remember that in this paper we are considering
couplings satisfying Eq. (\ref{eqQ}), that is

\begin{eqnarray}
\frac{1}{Q^{2}}\frac{dQ}{d\phi} & = & c.
\end{eqnarray}
This is equivalent to 

\begin{eqnarray}
\frac{d\log Q}{d\psi} & = & c.
\end{eqnarray}
This means the Horndeski Lagrangian can be written as 
\begin{eqnarray}
\mathcal{L}_{H} & = & X_{\psi}g(X_{\psi}e^{\lambda\psi})-[d_{1}X_{\psi}e^{\lambda\psi}+{\ell_{1}}X_{\psi}^{2}e^{2\lambda\psi}]\left(\Box\psi+2cX_{\psi}\right)\nonumber \\
 &  & +\biggl(h(\psi)+\frac{1}{2}d_{2}X_{\psi}^{2}e^{2\lambda\psi}+\frac{1}{3}{\ell_{2}}X_{\psi}^{3}e^{3\lambda\psi}\biggr)R\nonumber \\
 &  & +\biggl(d_{2}X_{\psi}e^{2\lambda\psi}+\ell_{2}X_{\psi}^{2}e^{3\lambda\psi}\biggr)\left[(\Box\psi)^{2}-\boxtimes\phi+4c\left(X_{\psi}\Box\psi+\frac{1}{2}\partial^{\mu}\psi\partial_{\nu}\psi\nabla_{\mu}\nabla^{\nu}\psi\right)\right]\nonumber \\
 &  & +qX_{\psi}^{2}e^{3\lambda\psi}G_{\mu\nu}\left(\nabla^{\mu}\nabla^{\nu}\psi-c\partial^{\mu}\psi\partial^{\nu}\psi\right)\nonumber \\
 &  & -\biggl(\frac{1}{6}\biggr)2qX_{\psi}e^{3\lambda\psi}\biggl\{(\Box\psi)^{3}-3(\Box\psi)\boxtimes\psi+2\boxdot\psi\nonumber \\
 &  & +c\biggl[6(\Box\psi)^{2}X_{\psi}-6X_{\psi}\boxtimes\psi+6(\Box\psi)\partial_{\mu}\psi\partial_{\nu}\psi\nabla^{\mu}\nabla^{\nu}\psi\nonumber \\
 &  & -2\nabla^{\mu}\nabla_{\alpha}\psi\nabla^{\alpha}\nabla_{\beta}\psi\partial^{\beta}\psi\partial_{\mu}\psi-2\partial^{\mu}\psi\partial_{\alpha}\psi\nabla^{\alpha}\nabla_{\beta}\psi\nabla^{\beta}\nabla{}_{\mu}\psi\nonumber \\
 &  & -2\nabla^{\mu}\nabla_{\alpha}\psi\partial^{\alpha}\psi\partial_{\beta}\psi\nabla^{\beta}\nabla_{\mu}\psi\biggr)\biggr]\nonumber \\
 &  & +c^{2}\biggl[12X_{\psi}\partial_{\mu}\psi\partial_{\nu}\psi\nabla^{\mu}\nabla^{\nu}\psi-2\nabla^{\mu}\nabla_{\alpha}\psi\partial^{\alpha}\psi\partial_{\beta}\psi\partial^{\beta}\psi\partial_{\mu}\psi\nonumber \\
 &  & -2\partial^{\mu}\psi\partial_{\alpha}\psi\nabla^{\alpha}\nabla_{\beta}\psi\partial^{\beta}\psi\partial_{\mu}\psi-2\partial^{\mu}\psi\partial_{\alpha}\psi\partial^{\alpha}\psi\partial_{\beta}\psi\nabla^{\beta}\nabla_{\mu}\psi\biggr]\nonumber \\
 &  & +c^{3}(-8X_{\psi}^{2})\biggr\}.\label{eq:Lpsi}
\end{eqnarray}
That is, when expressed in terms of $\psi$, there is no dependence
at all on the Lagrangian in the coupling (at least considering Eq.
(\ref{eqQ}) to be valid). This can be understood since the definition
of the coupling $Q$, given by Eq. (\ref{Qdef}), when expressed in
terms of $\psi$, assumes the form of a constant coupling:

\begin{eqnarray}
1 & = & -\frac{1}{\rho_{m}\sqrt{-g}}\frac{\delta S_{m}}{\delta\psi}.
\end{eqnarray}
This shows that, for the full Horndeski Lagrangian, the Lagrangian
with scaling solutions written in terms of $\phi$ and its derivatives,
with a general coupling satisfying $\frac{d\log Q}{d\phi}$ constant,
is equivalent to the above Lagrangian written in terms of $\psi$
and its derivatives, for the coupling $Q=1$. In other words, if one
is interested in scaling solutions with couplings $\frac{d\log Q}{d\phi}$
constant, it is sufficient to work with a Lagrangian given by Eq.
(\ref{eq:Lpsi}), independently of the particular coupling considered.
\textcolor{black}{It is important to remark, however, that the general
form of the full Horndeski Lagrangian found in this work do not apply
to more general couplings with $\frac{d\log Q}{d\phi}$ not constant.
Also, remember that the restriction for the couplings given by Eq.
(\ref{eqQ}) appears after Eq. (\ref{Boxphi}). That is, if one has
a Horndeski Lagrangian with $G_{3}=G_{5}=0$ and $G_{4}=1/2$, we
have a Lagrangian $\mathcal{L}_{H}(X,\phi)=X_{\psi}(X_{\psi}e^{\lambda\psi})+\frac{1}{2}R$
with no restrictions at all in the coupling $Q$.} In that case one
can indeed affirm, as done in Ref.\cite{Amendola:2006qi}, that the
case of constant coupling $Q=1$ is the most general. 

Now we make the redefinition $\bar{\phi}=\psi/\bar{Q},$ with $\bar{Q}$
a constant. This helps to compare with the literature, e.g. \cite{pt,ts,Amendola:2006qi,2014JCAP...03..041G}.
In the exponents that appear in the Lagrangian this is equivalent
to a redefinition of $\lambda$. Then, after dropping the bars we
can rewrite $\mathcal{L}_{H}$ as

\begin{eqnarray}
\mathcal{L}_{H} & = & Q^{2}Xg(Q^{2}Y)-[d_{1}Q^{2}Y+{\ell_{1}}Q^{4}Y^{2}]\left(Q\Box\phi+2(cQ)QX\right)+\biggl(h(Q\phi)+\frac{1}{2}d_{2}Q^{4}Y^{2}+\frac{1}{3}{\ell_{2}}Q^{6}Y^{3}\biggr)R\nonumber \\
\nonumber \\
 &  & +\biggl(d_{2}Q^{4}\frac{Y^{2}}{X}+\ell_{2}Q^{4}\frac{Y^{3}}{X}\biggr)\left[Q^{2}(\Box\phi)^{2}-Q^{2}\boxtimes\phi+4(cQ)\left(Q^{2}X\Box\phi+Q^{2}\frac{1}{2}\partial^{\mu}\phi\partial_{\nu}\phi\nabla_{\mu}\nabla^{\nu}\phi\right)\right]\nonumber \\
 &  & +qQ^{4}\frac{Y^{3}}{X}G_{\mu\nu}\left(Q\nabla^{\mu}\nabla^{\nu}\phi-(cQ)Q\partial^{\mu}\phi\partial^{\nu}\phi\right)\nonumber \\
 &  & -\biggl(\frac{1}{6}\biggr)2qQ^{2}\frac{Y^{3}}{X^{2}}\biggl\{ Q^{3}(\Box\phi)^{3}-3Q^{3}(\Box\phi)\boxtimes\phi+2Q^{3}\boxdot\phi\nonumber \\
 &  & +(cQ)\biggl[6Q^{3}(\Box\phi)^{2}X-6Q^{3}X\boxtimes\phi+6Q^{3}(\Box\phi)\partial_{\mu}\phi\partial_{\nu}\phi\nabla^{\mu}\nabla^{\nu}\phi\nonumber \\
 &  & -2Q^{3}\nabla^{\mu}\nabla_{\alpha}\phi\nabla^{\alpha}\nabla_{\beta}\phi\partial^{\beta}\phi\partial_{\mu}\phi-2Q^{3}\partial^{\mu}\phi\partial_{\alpha}\phi\nabla^{\alpha}\nabla_{\beta}\phi\nabla^{\beta}\nabla{}_{\mu}\phi\nonumber \\
 &  & -2Q^{3}\nabla^{\mu}\nabla_{\alpha}\phi\partial^{\alpha}\phi\partial_{\beta}\phi\nabla^{\beta}\nabla_{\mu}\phi\biggr)\biggr]\nonumber \\
 &  & +(cQ)^{2}\biggl[12Q^{3}X\partial_{\mu}\phi\partial_{\nu}\phi\nabla^{\mu}\nabla^{\nu}\phi-2Q^{3}\nabla^{\mu}\nabla_{\alpha}\phi\partial^{\alpha}\phi\partial_{\beta}\phi\partial^{\beta}\phi\partial_{\mu}\phi\nonumber \\
 &  & -2Q^{3}\partial^{\mu}\phi\partial_{\alpha}\phi\nabla^{\alpha}\nabla_{\beta}\phi\partial^{\beta}\phi\partial_{\mu}\phi-2Q^{3}\partial^{\mu}\phi\partial_{\alpha}\phi\partial^{\alpha}\phi\partial_{\beta}\phi\nabla^{\beta}\nabla_{\mu}\phi\biggr]\nonumber \\
 &  & +(cQ)^{3}(-8Q^{3}X^{2})\biggr\}.\label{eq:Lpsi-1}
\end{eqnarray}
with

\begin{eqnarray}
\lambda & = & -Q\left(\frac{1+w_{\mathrm{eff}}}{w_{\mathrm{eff}}}\right)\\
Y & = & Xe^{\lambda\psi}.
\end{eqnarray}
Since the system has symmetry under a simultaneous change of sign
of $\lambda$ and $Q$, we will be considering from now on $\lambda>0$.
Next, redefining the general constants and functions we get

\begin{eqnarray}
\mathcal{L}_{H} & = & Xg(Y)-[d_{1}Y+{\ell_{1}}Y^{2}]\left(\Box\phi+2cX\right)+\biggl(h(\phi)+\frac{1}{2}d_{2}Y^{2}+\frac{1}{3}{\ell_{2}}Y^{3}\biggr)R\nonumber \\
\nonumber \\
 &  & +\biggl(d_{2}\frac{Y^{2}}{X}+\ell_{2}\frac{Y^{3}}{X}\biggr)\left[(\Box\phi)^{2}-\boxtimes\phi+4c\left(X\Box\phi+\frac{1}{2}\partial^{\mu}\phi\partial_{\nu}\phi\nabla_{\mu}\nabla^{\nu}\phi\right)\right]\nonumber \\
 &  & +q\frac{Y^{3}}{X}G_{\mu\nu}\left(\nabla^{\mu}\nabla^{\nu}\phi-c\partial^{\mu}\phi\partial^{\nu}\phi\right)-\frac{q}{3}\frac{Y^{3}}{X^{2}}\biggl\{(\Box\phi)^{3}-3(\Box\phi)\boxtimes\phi+2\boxdot\phi\nonumber \\
 &  & +c\biggl[6(\Box\phi)^{2}X-6X\boxtimes\phi+6(\Box\phi)\partial_{\mu}\phi\partial_{\nu}\phi\nabla^{\mu}\nabla^{\nu}\phi-2\nabla^{\mu}\nabla_{\alpha}\phi\nabla^{\alpha}\nabla_{\beta}\phi\partial^{\beta}\phi\partial_{\mu}\phi-2\partial^{\mu}\phi\partial_{\alpha}\phi\nabla^{\alpha}\nabla_{\beta}\phi\nabla^{\beta}\nabla{}_{\mu}\phi\nonumber \\
 &  & -2\nabla^{\mu}\nabla_{\alpha}\phi\partial^{\alpha}\phi\partial_{\beta}\phi\nabla^{\beta}\nabla_{\mu}\phi\biggr)\biggr]+c{}^{2}\biggl[12X\partial_{\mu}\phi\partial_{\nu}\phi\nabla^{\mu}\nabla^{\nu}\phi-2\nabla^{\mu}\nabla_{\alpha}\phi\partial^{\alpha}\phi\partial_{\beta}\phi\partial^{\beta}\phi\partial_{\mu}\phi\nonumber \\
 &  & -2\partial^{\mu}\phi\partial_{\alpha}\phi\nabla^{\alpha}\nabla_{\beta}\phi\partial^{\beta}\phi\partial_{\mu}\phi-2\partial^{\mu}\phi\partial_{\alpha}\phi\partial^{\alpha}\phi\partial_{\beta}\phi\nabla^{\beta}\nabla_{\mu}\phi\biggr]-8c{}^{3}X^{2}\biggr\}.\label{eq:Lscal2}
\end{eqnarray}
\textcolor{black}{Now comparing the expressions for the Horndeski
Lagrangian given by Eqs. (\ref{L_scal}) and (\ref{eq:Lscal2}) one
sees that the latter expression is more complex than the former. The
two equivalent versions are related by field redefinitions: in the
form given by Eq. (\ref{L_scal}), the general Horndeski functions
$K,G_{3},G_{4},G_{5}$ are more evident, but the Lagrangian depends
on the coupling $Q$; the form given by Eq. (\ref{eq:Lscal2}) displays
a constant coupling, but has a more intrincate structure.}

\section{Conclusions}

This paper offers two insights into the vast realm of the Hornesdki
Lagrangians. First, we show that the entire Lagrangian is equivalent
to the scalar field pressure (at least in a FLRW metric), extending
earlier results valid for the $K,G_{3}$ subclass. This result is
then employed to identify the form of the Horndeski Lagrangian that
contains scaling solutions in which the ratio of matter to field density
and the equation of state are constant. This also generalizes previous
results, in particular Ref. \cite{2014JCAP...03..041G}. 

The existence of this particular class of solution is interesting
since it could represent a solution of the coincidence problem. If
the ratio $\Omega_{m}/\Omega_{\phi}$ depends on the fundamental constant
of the theory, instead of on initial conditions, then the fact that
it is close to unity would no longer be a surprising coincidence.
On the other hand, nothing guarantees that such a scaling solution
is stable or viable, in the sense of providing a valid cosmology when
compared to observations. Indeed we found in Ref. \cite{2014JCAP...03..041G}
that this was not the case for the subclass $K,G_{3}$ because one
has either a stable scaling accelerated regime or a well-behaved matter
era, but not both. We conjecture that this negative result extends
to the full scaling Horndeski Lagrangian \textcolor{black}{(\ref{L_scal})
or (\ref{eq:Lscal2})} but due to its very complicate form a full
proof is still to be elaborated.

\section{acknowledgements}

A. G. thanks FAPEMA and CNPq for financial support and ITP for the
hospitality. L.A. acknowledges support from DFG under the TransRegio33
project ``The Dark Universe''. L.A. acknowledges interesting discussions
with Ignacy Sawicki.

\section*{Appendix A - Some useful relations for flrw metric}

Metric determinant 

\begin{eqnarray}
\frac{d}{dt}\sqrt{-g} & = & \sqrt{-g}(3H).
\end{eqnarray}
Nonvanishing elements of the affine connection: 

\begin{eqnarray}
\Gamma_{ij}^{0} & = & A\dot{A}\eta_{ij,}\\
\Gamma_{0j}^{i} & =\Gamma_{j0}^{i}= & H\delta_{j}^{i},\\
\Gamma_{22}^{1} & = & -r,\\
\Gamma_{33}^{1} & = & -r\sin^{2}\theta,\\
\Gamma_{12}^{2} & = & \Gamma_{21}^{2}=\Gamma_{13}^{3}=\Gamma_{31}^{3}=\frac{1}{r},\\
\Gamma_{33}^{2} & = & -\sin\theta\cos\theta,\\
\Gamma_{23}^{3} & = & \Gamma_{32}^{3}=\cot\theta.
\end{eqnarray}
Nonvanishing components of Ricci tensor and Ricci scalar:

\begin{eqnarray}
R_{00} & = & -3\frac{\ddot{A}}{A},\\
R_{ij} & = & (A\ddot{A}+2\dot{A}^{2})\eta_{ij},\\
R & = & 6\dot{H}+12H^{2}.
\end{eqnarray}
Relation between $R$ and $p$, after using Einstein equations in
the above expression of $R$: 

\begin{eqnarray}
R & = & p\left(\frac{1}{w_{\phi}\Omega_{\phi}}-3\right).
\end{eqnarray}
Components of Einstein tensor:

\begin{eqnarray}
G_{\mu}^{\mu} & = & -(6\dot{H}+12H^{2}),\\
G_{0}^{0} & = & -3H^{2},\\
G_{j}^{i} & = & -(2\dot{H}+3H^{2})\delta_{j}^{i}.
\end{eqnarray}
We consider now the operators acting on $\phi$:

\begin{eqnarray}
\Box\phi & = & \nabla_{\mu}\nabla^{\mu}\phi=\frac{1}{\sqrt{-g}}\partial_{\mu}(\sqrt{-g}\nabla^{\mu}\phi)=-3H\dot{\phi}-\ddot{\phi},\\
\boxtimes\phi & = & \nabla_{\mu}\nabla_{\nu}\phi\nabla^{\mu}\nabla^{\nu}\phi=(\ddot{\phi})^{2}+3H^{2}(\dot{\phi})^{2},\\
\boxbar\phi & = & G_{\mu\nu}\nabla^{\mu}\nabla^{\nu}\phi=3H^{2}\ddot{\phi}+6H\dot{H}\dot{\phi}+9H^{3}\dot{\phi},\\
\boxdot\phi & = & (\nabla^{\mu}\nabla_{\alpha}\phi)(\nabla^{\alpha}\nabla_{\beta}\phi)(\nabla^{\beta}\nabla_{\mu}\phi)=-(\ddot{\phi})^{3}-3H^{3}(\dot{\phi})^{3}.
\end{eqnarray}

\section*{Appendix B - Some useful results for the master equation}

From Eq. (\ref{dphidN}) we have
\begin{eqnarray}
\dot{\phi} & = & -\frac{3H}{Q}w_{\phi}\Omega_{\phi},\\
\ddot{\phi} & = & -\frac{3\dot{H}}{Q}w_{\phi}\Omega_{\phi}-\frac{9H^{2}}{Q^{3}}\frac{dQ}{d\phi}(w_{\phi}\Omega_{\phi})^{2}.
\end{eqnarray}
Now, with 
\begin{eqnarray}
\dot{H} & = & -\frac{3}{2}H^{2}(1+w_{eff}),\\
H^{2} & = & \frac{1}{3}\frac{p}{w_{eff}},
\end{eqnarray}
we get

\begin{eqnarray}
\ddot{\phi} & = & \frac{3}{2}\frac{p}{Q}(1+w_{eff})\left(1+\frac{2}{\lambda}\frac{1}{Q^{2}}\frac{dQ}{d\phi}\right),\\
\nonumber \\
\nonumber 
\end{eqnarray}
with 

\begin{eqnarray}
\lambda & = & -\frac{1+w_{eff}}{w_{\phi}\Omega_{\phi}}.
\end{eqnarray}
Then the last expressions of Appendix A can be rewritten as 

\begin{eqnarray}
\Box\phi & = & \frac{3}{2}(1-w_{eff})\frac{p}{Q}\left(1-\frac{2}{\lambda}\frac{1+w_{eff}}{1-w_{eff}}\frac{1}{Q^{2}}\frac{dQ}{d\phi}\right),\\
\boxtimes\phi & = & \frac{p^{2}}{Q^{2}}\left\{ \left[\frac{3}{2}(1+w_{eff})\left(1+\frac{2}{\lambda}\frac{1}{Q^{2}}\frac{dQ}{d\phi}\right)\right]^{2}+3\right\} ,\\
\boxbar\phi & = & \frac{p^{2}}{Q}\left\{ \frac{1}{w_{eff}}\frac{3}{2}(1+w_{eff})\left(3+\frac{2}{\lambda}\frac{1}{Q^{2}}\frac{dQ}{d\phi}\right)-\frac{3}{w_{eff}}\right\} ,\\
\boxdot\phi & = & -\frac{p^{3}}{Q^{3}}\left\{ \left[\frac{3}{2}(1+w_{eff})\left(1+\frac{2}{\lambda}\frac{1}{Q^{2}}\frac{dQ}{d\phi}\right)\right]^{3}-3\right\} .
\end{eqnarray}
Now for $\frac{1}{Q^{2}}\frac{dQ}{d\phi}$ constant,

\begin{eqnarray}
\ln\Box\phi & = & \ln p-\ln Q+constant,\\
\ln\boxtimes\phi & = & 2\ln p-2\ln Q+constant,\\
\ln\boxbar\phi & = & 2\ln p-\ln Q+constant,\\
\ln\boxdot\phi & = & 3\ln p-3\ln Q+constant.
\end{eqnarray}
Then the corresponding partial derivatives with respect to $N$ are

\begin{eqnarray}
\frac{\partial\ln\Box\phi}{\partial N} & = & \frac{\partial\ln p}{\partial N}-\frac{\partial\ln Q}{\partial N}=-3(1+w_{eff})\left(1+\frac{1}{\lambda Q^{2}}\frac{dQ}{d\phi}\right),\\
\frac{\partial\ln\boxtimes\phi}{\partial N} & = & 2\frac{\partial\ln p}{\partial N}-2\frac{\partial\ln Q}{\partial N}=-3(1+w_{eff})\left(2+\frac{2}{\lambda Q^{2}}\frac{dQ}{d\phi}\right),\\
\frac{\partial\ln\boxbar\phi}{\partial N} & = & 2\frac{\partial\ln p}{\partial N}-\frac{\partial\ln Q}{\partial N}=-3(1+w_{eff})\left(2+\frac{1}{\lambda Q^{2}}\frac{dQ}{d\phi}\right),\\
\frac{\partial\ln\boxdot\phi}{\partial N} & = & 3\frac{\partial\ln p}{\partial N}-3\frac{\partial\ln Q}{\partial N}=-3(1+w_{eff})\left(3+\frac{3}{\lambda Q^{2}}\frac{dQ}{d\phi}\right).
\end{eqnarray}

\section*{Appendix C - Equality between lagrangian and pressure}

Here we prove in detail the equivalence of the Lagrangian and the
scalar field pressure in a FLRW metric. For completeness we rederive
the already known results for $K$ and $G_{3}$. The extension to
$G_{4},G_{5}$ is new.

\subsection{Term $K(\phi,X)$}

We start with 

\begin{eqnarray}
S_{total} & = & \int d^{4}x\frac{1}{2}R-\int d^{4}x\sqrt{-g}K(\phi,X)\\
 & = & S_{EH}+S,
\end{eqnarray}
That is, the Lagrangian in the action S is 

\begin{equation}
L=K(\phi,X).
\end{equation}
It is evident then that 

\begin{equation}
p=L=K.
\end{equation}

\subsection{Term $G_{3}$:}

We start with 

\begin{eqnarray}
S_{total} & = & \int d^{4}x\frac{1}{2}R-\int d^{4}x\sqrt{-g}G_{3}(\Box\phi)=\\
 & = & S_{EH}+S,
\end{eqnarray}
with 

\begin{eqnarray}
S & = & -\int d^{4}x\sqrt{-g}G_{3}(\Box\phi).\\
\nonumber 
\end{eqnarray}
That is, the Lagrangian in the action S is 

\begin{equation}
L=-G_{3}(\Box\phi)=-G_{3}(-3H\dot{\phi}-\ddot{\phi}).
\end{equation}
We have

\begin{equation}
p=-2X(G_{3,\phi}+\ddot{\phi}G_{3,X}).
\end{equation}
Now we compare with the Lagrangian 

\begin{eqnarray}
L & = & -G_{3}\Box\phi\nonumber \\
 & = & -\nabla_{\mu}(G_{3}\nabla^{\mu}\phi)+(\nabla_{\mu}G_{3})(\nabla^{\mu}\phi)\nonumber \\
 & = & -\nabla_{\mu}(G_{3}\nabla^{\mu}\phi)+p.
\end{eqnarray}

We note that the $L=p$, up to a covariant divergence, which integrated
in the action, by Gauss's law results to be zero.

\subsection{Term $G_{4}:$}

We start with 

\begin{eqnarray}
S_{total} & = & \int d^{4}x\frac{1}{2}R+\int d^{4}x\sqrt{-g}\left[\left(G_{4}-\frac{1}{2}\right)R+G_{4,X}[(\Box\phi)^{2}-(\boxtimes\phi)\right]\\
 & = & S_{EH}+S=I+III,
\end{eqnarray}
The Lagrangian in the action $S$ is 

\begin{equation}
L=\left(G_{4}-\frac{1}{2}\right)R+G_{4,X}[(\Box\phi)^{2}-(\boxtimes\phi)].
\end{equation}
We have 

\begin{eqnarray}
p & = & -(3H^{2}+2\dot{H})(1-2G_{4})-12H^{2}XG_{4,X}\nonumber \\
 &  & -4H\dot{X}G_{4,X}-8\dot{H}XG_{4,X}-8HX\dot{X}G_{4,XX}+2(\ddot{\phi}+2H\dot{\phi})G_{4,\phi}\nonumber \\
 &  & +4XG_{4,\phi\phi}+4X(\ddot{\phi}-2H\dot{\phi})G_{4,\phi X}.
\end{eqnarray}
Now we want to compare $p$ with the Lagrangian. We start with

\begin{eqnarray}
I & = & \int d^{4}x\sqrt{-g}\left(G_{4}-\frac{1}{2}\right)R\\
 & = & \int d^{4}x\sqrt{-g}\left(1-2G_{4}\right)(-3H^{2}-2\dot{H})+\int d^{4}x\sqrt{-g}\left(1-2G_{4}\right)(-3H^{2}-\dot{H})\nonumber \\
 & = & \int d^{4}x\sqrt{-g}\left(1-2G_{4}\right)(-3H^{2}-2\dot{H})+II+\int d^{4}x\sqrt{-g}\left(1-2G_{4}\right)(-\dot{H}),
\end{eqnarray}
where

\begin{eqnarray}
II & = & \int d^{4}x\sqrt{-g}\left(1-2G_{4}\right)(-3H^{2})\\
 & = & \int d^{4}x\left(\frac{d}{dt}\sqrt{-g}\right)\left(1-2G_{4}\right)(-H)\nonumber \\
 & = & -\int d^{4}x\sqrt{-g}\frac{d}{dt}\left[\left(1-2G_{4}\right)(-H)\right]\nonumber \\
 & = & -\int d^{4}x\sqrt{-g}\left[\left(-2G_{4,X}\dot{X}-2G_{4,\phi}\dot{\phi}\right)(-H)+\left(1-2G_{4}\right)(-\dot{H})\right].
\end{eqnarray}
Back to $I$:
\begin{eqnarray}
I & = & \int d^{4}x\sqrt{-g}\left(1-2G_{4}\right)(-3H^{2}-2\dot{H})\nonumber \\
 &  & +\int d^{4}x\sqrt{-g}\left[\left(2G_{4,X}\dot{X}+2G_{4,\phi}\dot{\phi}\right)(-H)-\left(1-2G_{4}\right)(-\dot{H})\right]+\int d^{4}x\sqrt{-g}\left(1-2G_{4}\right)(-\dot{H})\nonumber \\
 & = & \int d^{4}x\sqrt{-g}\left(1-2G_{4}\right)(-3H^{2}-2\dot{H})\nonumber \\
 &  & +\int d^{4}x\sqrt{-g}\left(-2H\dot{X}G_{4,X}-2H\dot{\phi}G_{4,\phi}\right).
\end{eqnarray}
Now consider the second part of the action $S$:

\begin{eqnarray}
III & = & \int d^{4}x\sqrt{-g}G_{4,X}[(\Box\phi)^{2}-(\boxtimes\phi)]\\
 & = & \int d^{4}x\sqrt{-g}G_{4,X}[(-3H\dot{\phi}-\ddot{\phi})^{2}-((\ddot{\phi})^{2}+3H^{2}(\dot{\phi})^{2})]\nonumber \\
 & = & \int d^{4}x\sqrt{-g}G_{4,X}[6H^{2}(\dot{\phi})^{2}+6H\dot{\phi}\ddot{\phi}]\nonumber \\
 & = & \int d^{4}x\sqrt{-g}G_{4,X}[6H^{2}(\dot{\phi})^{2}]+IV,
\end{eqnarray}
where

\begin{eqnarray}
IV & = & \int d^{4}x\sqrt{-g}G_{4,X}[6H\dot{\phi}\ddot{\phi}]\\
 & = & 6\int d^{4}x\sqrt{-g}G_{4,X}H\dot{\phi}\left(\frac{d}{dt}\dot{\phi}\right)\nonumber \\
 & = & -6\int d^{4}x\frac{d}{dt}\left(\sqrt{-g}G_{4,X}H\dot{\phi}\right)\dot{\phi}\nonumber \\
 & = & -6\int d^{4}x\sqrt{-g}\left(3HG_{4,X}H\dot{\phi}+(G_{4,XX}\dot{X}+G_{4,X\phi}\dot{\phi})H\dot{\phi}+G_{4,X}\dot{H}\dot{\phi}+G_{4,X}H\ddot{\phi}\right)\dot{\phi}\nonumber \\
 & = & \int d^{4}x\sqrt{-g}\left[\left(-36H^{2}X-12\dot{H}X-6H\dot{X}\right)G_{4,X}-12X\dot{X}HG_{4,XX}-12G_{4,X\phi}HX\dot{\phi}\right].
\end{eqnarray}
Then back to $III$:

\begin{eqnarray}
III & = & \int d^{4}x\sqrt{-g}G_{4,X}[(\Box\phi)^{2}-(\boxtimes\phi)]\nonumber \\
 & = & \int d^{4}x\sqrt{-g}G_{4,X}[6H^{2}(\dot{\phi})^{2}]+IV\nonumber \\
 & = & \int d^{4}x\sqrt{-g}\left[\left(-24H^{2}X-6H\dot{X}-12\dot{H}X\right)G_{4,X}-12X\dot{X}HG_{4,XX}-12HX\dot{\phi}G_{4,X\phi}\right].
\end{eqnarray}
Now from $I$ and $III$ we recover the action:

\begin{eqnarray}
S & = & \int d^{4}x\sqrt{-g}\left[\left(G_{4}-\frac{1}{2}\right)R+G_{4,X}[(\Box\phi)^{2}-(\boxtimes\phi)\right]=I+III\\
 & = & \int d^{4}x\sqrt{-g}\left(1-2G_{4}\right)(-3H^{2}-2\dot{H})+\int d^{4}x\sqrt{-g}\left(-2H\dot{X}G_{4,X}-2H\dot{\phi}G_{4,\phi}\right)\nonumber \\
 &  & +\int d^{4}x\sqrt{-g}\left[\left(-24H^{2}X-6H\dot{X}-12\dot{H}X\right)G_{4,X}-12X\dot{X}HG_{4,XX}-12HX\dot{\phi}G_{4,X\phi}\right]\nonumber \\
 & = & \int d^{4}x\sqrt{-g}p+\int d^{4}x\sqrt{-g}\{[(3H)(-4H^{2}X)+\frac{d}{dt}(-4HX)]G_{4,X}\nonumber \\
 &  & +(-4XH)(\dot{X}G_{4,XX}+\dot{\phi}G_{4,X\phi})\}\nonumber \\
 &  & +\int d^{4}x\sqrt{-g}\{(-2\ddot{\phi}-6H\dot{\phi})G_{4,\phi}-4XG_{4,\phi\phi}+(-4X\ddot{\phi})G_{4,X\phi})\}\nonumber \\
 & = & \int d^{4}x\sqrt{-g}p+\int d^{4}x\frac{d}{dt}\left(\sqrt{-g}(-4H^{2}X)G_{4,X}\right)\nonumber \\
 &  & +\int d^{4}x\sqrt{-g}\{(-2\ddot{\phi}-6H\dot{\phi})G_{4,\phi}-2\dot{\phi}\dot{\phi}G_{4,\phi\phi}-2\dot{\phi}\dot{X})G_{4,X\phi})\}\nonumber \\
 & = & \int d^{4}x\sqrt{-g}p+\int d^{4}x\frac{d}{dt}\left(\sqrt{-g}(-4H^{2}X)G_{4,X}\right)\nonumber \\
 &  & +\int d^{4}x\{-2\sqrt{-g}\left(\frac{d}{dt}(\dot{\phi})\right)G_{4,\phi}-2\left(\frac{d}{dt}(\sqrt{-g})\right)(\dot{\phi}G_{4,\phi})-2\sqrt{-g}\dot{\phi}\frac{d}{dt}\left(G_{4,\phi}\right)\}\nonumber \\
 & = & \int d^{4}x\sqrt{-g}p+\int d^{4}x\frac{d}{dt}\left(\sqrt{-g}(-4H^{2}X)G_{4,X}\right)\nonumber \\
 &  & +\int d^{4}x\frac{d}{dt}\left(-2\sqrt{-g}(\dot{\phi})G_{4,\phi}\right)\nonumber \\
 & = & \int d^{4}x\sqrt{-g}p.
\end{eqnarray}
That is again $L=p$.

\subsection{Term $G_{5}$:}

We start now with 

\begin{eqnarray}
S_{total} & = & \int d^{4}x\frac{1}{2}R+\int d^{4}x\sqrt{-g}\{G_{5}(\boxbar\phi)-\frac{1}{6}G_{5,X}[(\Box\phi)^{3}-3(\Box\phi)(\boxtimes\phi)+2(\boxdot\phi)]\}\\
 & = & S_{EH}+S
\end{eqnarray}
with

\begin{eqnarray}
S & = & \int d^{4}x\sqrt{-g}\{G_{5}(\boxbar\phi)-\frac{1}{6}G_{5,X}[(\Box\phi)^{3}-3(\Box\phi)(\boxtimes\phi)+2(\boxdot\phi)]\}.
\end{eqnarray}
That is, the Lagrangian in the action $S$ is 

\begin{equation}
L=G_{5}(\boxbar\phi)-\frac{1}{6}G_{5,X}[(\Box\phi)^{3}-3(\Box\phi)(\boxtimes\phi)+2(\boxdot\phi)].
\end{equation}
We have 

\begin{eqnarray}
p & = & -2X(2H^{3}\dot{\phi}+2H\dot{H}\dot{\phi}+3H^{2}\ddot{\phi})G_{5,X}-4H^{2}X^{2}\ddot{\phi}G_{5,XX}\nonumber \\
 &  & +4HX(\dot{X}-HX)G_{5,\phi X}+2[2(\dot{H}X+H\dot{X})+3H^{2}X]G_{5,\phi}+4HX\dot{\phi}G_{5,\phi\phi}.
\end{eqnarray}
Now we want to compare $p$ with the Lagrangian:

\begin{eqnarray}
L & = & G_{5}(3H^{2}\ddot{\phi}+6H\dot{H}\dot{\phi}+9H^{3}\dot{\phi})\nonumber \\
 &  & -\frac{1}{6}G_{5,X}[(-3H\dot{\phi}-\ddot{\phi})^{3}-3(-3H\dot{\phi}-\ddot{\phi})((\ddot{\phi})^{2}+3H^{2}(\dot{\phi})^{2})+2(-(\ddot{\phi})^{3}-3H^{3}(\dot{\phi})^{3})]\\
 & = & G_{5}(3H^{2}\ddot{\phi}+6H\dot{H}\dot{\phi}+9H^{3}\dot{\phi})-\frac{1}{6}G_{5,X}\{[-27H^{3}(\dot{\phi})^{3}-(\ddot{\phi})^{3}+3(-3H\dot{\phi})^{2}(-\ddot{\phi})+3(-3H\dot{\phi})(-\ddot{\phi})^{2}]\nonumber \\
 &  & +[3(3H\dot{\phi}+\ddot{\phi})((\ddot{\phi})^{2}+6H^{2}X)]-2[(\ddot{\phi})^{3}+6H^{3}X(\dot{\phi})]\}\nonumber \\
 & = & G_{5}(3H^{2}\ddot{\phi}+6H\dot{H}\dot{\phi}+9H^{3}\dot{\phi})+G_{5,X}(6H^{2}X\ddot{\phi}+2H^{3}X\dot{\phi}).\\
\nonumber 
\end{eqnarray}
Now

\begin{eqnarray}
V & = & \int d^{4}x\sqrt{-g}G_{5,X}6H^{2}X\ddot{\phi}\\
 & = & -\int d^{4}x\frac{d}{dt}\left(\sqrt{-g}G_{5,X}6H^{2}X\right)\dot{\phi}\nonumber \\
 & = & -\int d^{4}x\biggl[\frac{d}{dt}\left(\sqrt{-g}\right)G_{5,X}6H^{2}X+\sqrt{-g}\frac{d}{dt}\left(G_{5,X}\right)6H^{2}X\nonumber \\
 &  & +\sqrt{-g}G_{5,X}\frac{d}{dt}\left(6H^{2}\right)X+\sqrt{-g}G_{5,X}6H^{2}\frac{d}{dt}\left(X\right)\biggr]\dot{\phi}\\
 & = & -\int d^{4}x\sqrt{-g}\left[3HG_{5,X}6H^{2}X+(G_{5,XX}\dot{X}+G_{5,X\phi}\dot{\phi})6H^{2}X+G_{5,X}\left(12H\dot{H}\right)X+G_{5,X}6H^{2}\dot{X}\right]\dot{\phi}\nonumber \\
 & = & \int d^{4}x\sqrt{-g}\left[-18H^{3}X\dot{\phi}G_{5,X}-12H^{2}X^{2}\ddot{\phi}G_{5,XX}-12H^{2}X^{2}G_{5,X\phi}-12H\dot{H}X\dot{\phi}G_{5,X}-12H^{2}X\ddot{\phi}G_{5,X}\right].
\end{eqnarray}
Then, the terms of the Lagrangian depending explicitly on $G_{5,X}$
can be written as

\begin{eqnarray}
 &  & \int d^{4}x\sqrt{-g}G_{5,X}(6H^{2}X\ddot{\phi}+2H^{3}X\dot{\phi})\\
 & = & \int d^{4}x\sqrt{-g}\biggl[G_{5,X}(-18H^{3}X\dot{\phi}-12H\dot{H}X\dot{\phi}-12H^{2}X\ddot{\phi}+2H^{3}X\dot{\phi})+G_{5,XX}(-12H^{2}X^{2}\ddot{\phi})\biggr]\nonumber \\
 &  & +G_{5,X\phi}(-12H^{2}X^{2})\\
 & = & \int d^{4}x\sqrt{-g}\left[G_{5,X}(-16H^{3}X\dot{\phi}-12H\dot{H}X\dot{\phi}-12H^{2}X\ddot{\phi})+G_{5,XX}(-12H^{2}X^{2}\ddot{\phi})+G_{5,X\phi}(-12H^{2}X^{2})\right].
\end{eqnarray}
Now 

\begin{eqnarray}
VI & = & \int d^{4}x\sqrt{-g}G_{5}3H^{2}\ddot{\phi}\\
 & = & 3\int d^{4}x\frac{d}{dt}\left(\sqrt{-g}G_{5}H^{2}\dot{\phi}\right)-3\int d^{4}x\frac{d}{dt}\left(\sqrt{-g}G_{5}H^{2}\right)\dot{\phi}\nonumber \\
 & = & -3\int d^{4}x\frac{d}{dt}\left(\sqrt{-g}G_{5}H^{2}\right)\dot{\phi}\nonumber \\
 & = & -3\int d^{4}x\left[\frac{d}{dt}\left(\sqrt{-g}\right)G_{5}H^{2}\dot{\phi}+\sqrt{-g}\frac{d}{dt}\left(G_{5}\right)H^{2}\dot{\phi}+\sqrt{-g}G_{5}\frac{d}{dt}\left(H^{2}\right)\dot{\phi}\right]\nonumber \\
 & = & \int d^{4}x\sqrt{-g}\left[-9H^{3}G_{5}\dot{\phi}-3G_{5,X}\dot{X}H^{2}\dot{\phi}-3G_{5,\phi}\dot{\phi}H^{2}\dot{\phi}-6G_{5}H\dot{H}\dot{\phi}\right].
\end{eqnarray}
Then, from the terms of the Lagrangian depending on $G_{5}$:

\begin{eqnarray}
 &  & \int d^{4}x\sqrt{-g}[(3H^{2}\ddot{\phi}+6H\dot{H}\dot{\phi}+9H^{3}\dot{\phi})G_{5}]\\
 & = & \int d^{4}x\sqrt{-g}\left[-9H^{3}G_{5}\dot{\phi}-3G_{5,X}\dot{X}H^{2}\dot{\phi}-3G_{5,\phi}\dot{\phi}H^{2}\dot{\phi}-6G_{5}H\dot{H}\dot{\phi}+(6H\dot{H}\dot{\phi}+9H^{3}\dot{\phi})G_{5}\right]\nonumber \\
 & = & \int d^{4}x\sqrt{-g}\left[-6G_{5,X}XH^{2}\ddot{\phi}-6G_{5,\phi}XH^{2}\right].
\end{eqnarray}
Then the action can be written as 

\begin{eqnarray}
S & = & \int d^{4}x\sqrt{-g}L\nonumber \\
 & = & \int d^{4}x\sqrt{-g}[G_{5}(3H^{2}\ddot{\phi}+6H\dot{H}\dot{\phi}+9H^{3}\dot{\phi})+G_{5,X}(6H^{2}X\ddot{\phi}+2H^{3}X\dot{\phi})]\nonumber \\
 & = & \int d^{4}x\sqrt{-g}\{[-6G_{5,X}XH^{2}\ddot{\phi}-6G_{5,\phi}XH^{2}]\nonumber \\
 &  & +[G_{5,X}(-16H^{3}X\dot{\phi}-12H\dot{H}X\dot{\phi}-12H^{2}X\ddot{\phi})+G_{5,XX}(-12H^{2}X^{2}\ddot{\phi})+G_{5,X\phi}(-12H^{2}X^{2})]\}\nonumber \\
 & = & \int d^{4}x\sqrt{-g}\{G_{5,X}(-16H^{3}X\dot{\phi}-12H\dot{H}X\dot{\phi}-18H^{2}X\ddot{\phi})\nonumber \\
 &  & +G_{5,XX}(-12H^{2}X^{2}\ddot{\phi})+G_{5,X\phi}(-12H^{2}X^{2})+(6XH^{2})G_{5,\phi}-(12XH^{2})G_{5,\phi}\}.
\end{eqnarray}
Now we transform the term of the action depending on $G_{5,\phi}:$

\begin{eqnarray}
-\int d^{4}x\sqrt{-g}(12XH^{2})G_{5,\phi} & = & -\int d^{4}x(3H\sqrt{-g})(4HX)G_{5,\phi}\nonumber \\
 & = & -\int d^{4}x\frac{d}{dt}\left[(\sqrt{-g})(4HXG_{5,\phi})\right]+\int d^{4}x\sqrt{-g}\frac{d}{dt}\left[4HXG_{5,\phi}\right]\nonumber \\
 & = & \int d^{4}x\sqrt{-g}\frac{d}{dt}\left[4HXG_{5,\phi}\right]\nonumber \\
 & = & \int d^{4}x\sqrt{-g}\left[(4\dot{H}X+4H\dot{X})G_{5,\phi}+(4HX\dot{X})G_{5,\phi X}+(4HX\dot{\phi})G_{5,\phi\phi}\right].
\end{eqnarray}
Then the action is

\begin{eqnarray}
S & = & \int d^{4}x\sqrt{-g}\{G_{5,X}(-16H^{3}X\dot{\phi}-12H\dot{H}X\dot{\phi}-18H^{2}X\ddot{\phi})\nonumber \\
 &  & +G_{5,XX}(-12H^{2}X^{2}\ddot{\phi})+G_{5,X\phi}(-12H^{2}X^{2})+(6XH^{2})G_{5,\phi}-(12XH^{2})G_{5,\phi}\}\nonumber \\
 & = & \int d^{4}x\sqrt{-g}\{G_{5,X}(-4H^{3}X\dot{\phi}-4H\dot{H}X\dot{\phi}-6H^{2}X\ddot{\phi})+(6XH^{2}+4\dot{H}X+4H\dot{X})G_{5,\phi}\nonumber \\
 &  & +(4HX\dot{\phi})G_{5,\phi\phi}+G_{5,XX}(-4H^{2}X^{2}\ddot{\phi})+G_{5,X\phi}(4HX\dot{\phi}-4H^{2}X^{2})\}\nonumber \\
 &  & +\int d^{4}x\sqrt{-g}\{G_{5,X}(-12H^{3}X\dot{\phi}-8H\dot{H}X\dot{\phi}-12H^{2}X\ddot{\phi})+G_{5,XX}(-8H^{2}X^{2}\ddot{\phi})+G_{5,X\phi}(-8H^{2}X^{2})\}\\
 & = & \int d^{4}x\sqrt{-g}p+S',
\end{eqnarray}
where

\begin{eqnarray}
S' & = & \int d^{4}x\sqrt{-g}\{G_{5,X}(-12H^{3}X\dot{\phi}-8H\dot{H}X\dot{\phi}-12H^{2}X\ddot{\phi})+G_{5,XX}(-8H^{2}X^{2}\ddot{\phi})+G_{5,X\phi}(-8H^{2}X^{2})\}\\
 & = & \int d^{4}x\sqrt{-g}\{G_{5,X}(-12H^{3}X\dot{\phi}-8H\dot{H}X\dot{\phi}-12H^{2}X\ddot{\phi})+\left(\frac{d}{dt}G_{5,X}\right)(-4H^{2}X\dot{\phi})\}\nonumber \\
 & = & \int d^{4}x\{G_{5,X}\sqrt{-g}[-12H^{3}X\dot{\phi}-8H\dot{H}X\dot{\phi}+(-4H^{2}\dot{\phi}\ddot{\phi}\dot{\phi}-4H^{2}X\dot{\phi})]\nonumber \\
 &  & +\sqrt{-g}\left(\frac{d}{dt}G_{5,X}\right)(-4H^{2}X\dot{\phi})\}\nonumber \\
 & = & \int d^{4}x\{G_{5,X}\frac{d}{dt}\left[\sqrt{-g}(-4H^{2}X\dot{\phi)}\right]+\sqrt{-g}\left(\frac{d}{dt}G_{5,X}\right)(-4H^{2}X\dot{\phi})\}\nonumber \\
 & = & \int d^{4}x\frac{d}{dt}\left[\sqrt{-g}G_{5,X}(-4H^{2}X\dot{\phi)}\right]=0.
\end{eqnarray}
That is, again $L=p$. This completes our proof.

\begin{thebibliography}{99}
\bibitem{Perlmutter_etal_1999}S. Perlmutter et al. (Supernova Cosmology Project), Astrophys. J. 517, 565 (1999), astro-ph/9812133.
\bibitem{Riess_etal_1998}A. G. Riess et al. (Supernova Search Team), Astron. J. 116, 1009 (1998), astro-ph/9805201.
\bibitem{Wetterich_1988}C. Wetterich, Nucl. Phys. B302, 668 (1988).
\bibitem{Ratra_Peebles_1988}B. Ratra and P. J. E. Peebles, Phys. Rev. D37, 3406 (1988).
\bibitem{Copeland:1997et}E. J. Copeland, A. R. Liddle, and D. Wands, Phys.Rev. D57, 4686 (1998), gr-qc/9711068.
\bibitem{Ferreira_Joyce_1998}P. G. Ferreira and M. Joyce, Phys. Rev. D58, 023503 (1998), astro-ph/9711102.
\bibitem{1998PhRvL..80.1582C}R. R. Caldwell, R. Dave, and P. J. Steinhardt, Physical Review Letters 80, 1582 (1998), astro-ph/9708069.
\bibitem{Wetterich_1995}C. Wetterich, Astron. Astrophys. 301, 321 (1995), hep-th/9408025.
\bibitem{amendola2000}L. Amendola, Phys.Rev. D62, 043511 (2000), astro-ph/9908023.
\bibitem{Baccigalupi_Matarrese_Perrotta_2000}C. Baccigalupi, S. Matarrese, and F. Perrotta, Phys. Rev. D62, 123510 (2000), astro-ph/0005543.
\bibitem{2006JCAP...12..020A}L. Amendola, C. Charmousis, and S. C. Davis, JCAP 12, 020 (2006), arXiv:hep-th/0506137.
\bibitem{kessence}C. Armendariz-Picon, V. F. Mukhanov, and P. J. Steinhardt, Phys. Rev. D63, 103510 (2001), astro-ph/0006373.
\bibitem{Horndeski:1974}G. W. Horndeski, Int.J.Th.Phys. 10, 363 (1974).
\bibitem{Deffayet:2011gz}C. Deffayet, X. Gao, D. Steer, and G. Zahariade, Phys.Rev. D84, 064039 (2011), 1103.3260.
\bibitem{Kobayashi:2011nu}T. Kobayashi, M. Yamaguchi, and J. Yokoyama, Prog.Theor.Phys. 126, 511 (2011), 1105.5723.
\bibitem{Gleyzes:2013ooa}J. Gleyzes, D. Langlois, F. Piazza, and F. Vernizzi (2013), 1304.4840.
\bibitem{2014PhRvD..89f4046Z}M. Zumalac\'arregui and J. Garc\'ia-Bellido, Phys. Rev. D 89, 064046 (2014), 1308.4685.
\bibitem{2012PhRvD..85j4040C}C. Charmousis, E. J. Copeland, A. Padilla, and P. M. Saffin, Phys. Rev. D 85, 104040 (2012), 1112.4866.
\bibitem{2014JCAP...03..041G}A. R. Gomes and L. Amendola, JCAP 3, 041 (2014), 1306.3593.
\bibitem{Kscal}L. Amendola, M. Quartin, S. Tsujikawa, and I. Waga, Phys.Rev. D74, 023525 (2006), astro-ph/0605488.
\bibitem{Amendola_Baldi_Wetterich_2008}L. Amendola, M. Baldi, and C. Wetterich, Phys. Rev. D 78, 023015 (2008), 0706.3064.
\bibitem{Amendola:2001rc}L. Amendola and D. Tocchini-Valentini, Phys.Rev. D66, 043528 (2002), astro-ph/0111535.
\bibitem{Baldi_2012a}M. Baldi, Annalen der Physik 524, 602 (2012), 1204.0514.
\bibitem{2015arXiv150205922V}V. Vardanyan and L. Amendola, ArXiv e-prints (2015), 1502.05922.
\bibitem{pt}F. Piazza and S. Tsujikawa, JCAP 0407, 004 (2004), hep-th/0405054.
\bibitem{ts}S. Tsujikawa and M. Sami, Phys.Lett. B603, 113 (2004), hep-th/0409212.
\bibitem{Amendola:2006qi}L. Amendola, M. Quartin, S. Tsujikawa, and I. Waga, Phys.Rev. D74, 023525 (2006), astro-ph/0605488.
\bibitem{Pujolas:2011he}O. Pujolas, I. Sawicki, and A. Vikman, JHEP 1111, 156 (2011), 1103.5360.
\bibitem{Deffayet:2010qz}C. Deffayet, O. Pujolas, I. Sawicki, and A. Vikman, JCAP 1010, 026 (2010), 1008.0048.
\bibitem{2000PhRvD..62b3511C}T. Chiba, T. Okabe, and M. Yamaguchi, Phys. Rev. D 62, 023511 (2000), astro-ph/9912463.
\bibitem{ArmendarizPicon:2000dh}C. Armendariz-Picon, V. F. Mukhanov, and P. J. Steinhardt, Phys.Rev.Lett. 85, 4438 (2000), astro-ph/0004134.
\bibitem{DeFelice:2011bh}A. De Felice and S. Tsujikawa, JCAP 1202, 007 (2012), 1110.3878.
\bibitem{tsujikawa_DMDE}S. Tsujikawa, Dark Energy: Investigation and Modeling, in Dark Matter and Dark Energy - A Challenge for Modern
Cosmology, Springer (2011).
\end{thebibliography}
\end{document}